\documentclass[runningheads]{llncs}
\usepackage{amsfonts}
\usepackage{hyperref}
\usepackage{graphicx}

\usepackage{floatrow}
\DeclareFloatFont{tiny}{\scriptsize}
\usepackage{comment}
\usepackage{siunitx}

\newcommand{\gitlink}[0]{\url{https://github.com/DocPierro/modelchecking_spd.git}}

\usepackage{algpseudocode,algorithm,algorithmicx}
\usepackage{multirow}
\usepackage{booktabs}
\usepackage{amsmath}
\usepackage{array}
\usepackage{tikz}
\usetikzlibrary{automata,shapes.multipart} 
\usetikzlibrary{arrows,petri}
\usetikzlibrary{arrows}
\usetikzlibrary{shapes,backgrounds}
\usetikzlibrary{arrows,shapes,backgrounds,patterns,decorations.pathreplacing,decorations.pathmorphing,decorations.markings,shadows,shapes.misc,calc,positioning}
\usetikzlibrary {graphs}

\newcommand{\flow}{\textit{flow}}

\newcommand{\cosmos}{\texttt{Cosmos}}

\newcommand{\veccyln}[2][{[n,m]}]{C({#2}, \eltS_{0:{#2}}, {#1}^{#2})}

\newcommand{\aut}{\mathcal{A}}

\usepackage[colorinlistoftodos]{todonotes}
\definecolor{forestgreen}{rgb}{0.13,0.55,0.13}
\definecolor{amber}{rgb}{1.0,0.49,0}
\definecolor{cargreen}{rgb}{0.0,0.8,0.6}

\usepackage[T1]{fontenc}
\usepackage{graphicx}
\usepackage{subcaption}

\usepackage{pgfplots}
\usepackage{subcaption}

\begin{document}
%
\title{Statistical  process discovery}

%
%
\author{ 
Pierre Cry\inst{1}
\and
Paolo Ballarini\inst{1}
\and
Andr\'as Horv\'ath\inst{2}
\and
Pascale Le Gall\inst{1}
}
%
%
\institute{Université Paris Saclay, CentraleSupélec, MICS, Gif-sur-Yvette, France 
\email{\{paolo.ballarini,pierre.cry,pascale.legall\}@centralesupelec.fr}\\
\and 
Università di Torino, Torino, Italy,
\email{horvath@di.unito.it}}
%
\maketitle              
\begin{abstract}

Stochastic process discovery is concerned with deriving a  model capable of reproducing the stochastic character of observed executions of a given process, stored in a log. This leads to an optimisation problem in which the model's parameter space is searched for,  driven by the resemblance between  the log's and the model's stochastic languages. The bottleneck of such optimisation problem lay  in  the determination of the model's stochastic language which existing approaches deal with through, hardly scalable, exact computation approaches. In this paper we introduce  a novel  framework in which we combine a simulation-based Bayesian parameter inference scheme, used to search for the  ``optimal'' instance of a stochastic  model, with an expressive statistical model checking  engine,   used (during inference) to approximate the language of the considered model's instance. 
Because of its simulation-based nature, the  payoff is  that, the runtime for discovering of the optimal instance of a model  can be easily traded in for accuracy, hence allowing to treat large models which would result in a prohibitive runtime  with non-simulation based alternatives. We validate our approach on several popular event logs concerning real-life systems.


\end{abstract}

\section{Introduction}
\label{sec:intro}

\noindent
{\it The process mining problem.}
The primary goal of \emph{process mining}~\cite{10.5555/2948762}  is    \emph{discovering} of formal models that adequately mimic the dynamics of a business process. Discovery relies on observations of the considered process stored as \emph{traces},  in an  \emph{event log}.  A \emph{trace} consists of a sequence of activities that represent one observed execution of the process. 
Classic process discovery  algorithms~\cite{DBLP:conf/apn/LeemansFA13,1316839,tonbeta166,10.1007/978-3-540-68746-7_24,VANDENBROUCKE2017109,10.1007/s10115-018-1214-x}, aim at capturing the workflow aspects of the considered process, that is,   extracting a so-called \emph{workflow} model, normally in the form of a structured Petri net, capable of reproducing the \emph{log's language}.
The quality of discovery algorithms is assessed by means of  \emph{conformance checking}  indicators such as, e.g., \emph{fitness}, which measures how much of the language (i.e., the set of unique traces) of the log is reproduced by the discovered model, and \emph{precision}, which, conversely,  quantifies how much of the language of the model (the traces issued by the model) is contained in the log. 

\noindent
{\it The stochastic process mining problem.} 
By taking into account  how often each unique sequence of actions has been detected while observing the process (i.e. trace multiplicity) leads to the stochastic  extension of the process discovery problem, whose aim is  to devise a stochastic model whose  \emph{stochastic language}  resembles that of the log. The resemblance is assessed via \emph{stochastic conformance} indicators, such as those based on  adaptations of the Earth Movers Distance (EMD)~\cite{10.1007/978-3-030-26643-1_8,LEEMANS2021101724} or those based on entropy  measures~\cite{ALKHAMMASH2022101922}.
Within the still relatively emergent literature~\cite{5f0e8dd04572478fb450eefe4211d69f,DBLP:journals/corr/abs-2406-10817,DBLP:conf/icpm/BurkeLW20,DBLP:conf/apn/BurkeLW21,10.1007/11494744_25,horvath2025probabilisticprocessdiscoverystochastic}, mining of a stochastic model is mainly achieved \emph{indirectly}, that is, first a non-stochastic, workflow model is extracted from the log through a standard discovery algorithm (hence disregarding trace multiplicity),  then the obtained model is converted into a stochastic one by associating \emph{weight} parameters to each event. 
This leads to a \emph{parameter optimization} problem whose goal is to identify adequate parameter (weight) values so that the corresponding stochastic language is \emph{as close as possible} to that of the log.

\noindent
{\it Our contribution.}  
Discovery of optimal parameters for a stochastic model requires an effective procedure to evaluate the stochastic language emitted by the model. 
Existing approaches~\cite{DBLP:journals/corr/abs-2406-10817,10.1007/978-3-031-61057-8_3} 
rely on exact computation of a model's stochastic language and,  although effective, suffer  from poor scalability, making it impractical for complex logs. 
As a remedy to such bottleneck, in this paper, we propose an alternative optimization  framework,  which relies on a \emph{simulation-based} engine  to obtain an (arbitrarily precise) approximation of the  model's stochastic language,   through a statistical model checking tool~\cite{BALLARINI201553}. 
The search for optimal model's parameters is then achieved through an adaptation of the Approximate Bayesian Computation (ABC)~\cite{Sisson2018,Marin2011} scheme, a \emph{likelihood-free}  Bayesian parameter inference method, in which the model's language approximation engine is plugged in. 
Experimental evidence shows that depending on the complexity of the considered event log (hence of the mined model), the simulation-based discovery of optimal stochastic models we propose here may be better than numerical-based ones, resulting in the discovery of more conformant models in less time.  


The paper is organized as follows: Section~\ref{sec:prelim} introduces preliminary notions used throughout the manuscript and succinctly overview background material including the HASL model checking framework and the ABC parameter inference scheme. 
In Section~\ref{sec:method}, the novel simulation-based framework for discovery of optimal instances of sWN models. 
In Section~\ref{sec:results}, we demonstrate the novel framework through several experiments on popular real-life event logs. We wrap up the manuscript with conclusive remarks and future perspectives in Section~\ref{sec:conclusion}. 


\noindent\emph{Related work.}
Several contributions have been proposed within the still thin yet growing literature on stochastic process discovery. 
In their seminal work  ~\cite{5f0e8dd04572478fb450eefe4211d69f} authors introduced a framework to discover generalized stochastic Petri net (GSPN) models extended with generally distributed timed transitions so to allow for performance analysis of the mined process. 
In close relationship with the problem we face in this paper, i.e. discovery of untimed, probabilistic models, Burke \emph{et al.}'s~\cite{DBLP:conf/icpm/BurkeLW20} instead addressed the problem of converting a workflow Petri net, mined through a conventional discovery algorithm, into an adequate (untimed) stochastic workflow Petri net through weight estimation. 
Specifically, \cite{DBLP:conf/icpm/BurkeLW20} introduced six weight estimators that combine summary statistics computed on the log with statistics computed on the model while also taking into account structural relationships between the Petri net nodes (e.g.,  \emph{transitions causality}). These estimators enjoy being computationally light as, by definition, they do not need to assess the language of the Petri net model. However, the price for such simplicity is paid in terms of conformance, as the distance between the resulting model's and the log's stochastic languages appears to be far from optimal in many cases. In a follow-up work~\cite{DBLP:conf/apn/BurkeLW21}, the same authors introduced a framework to directly discover an \emph{untimed} GSPN model from a log based on traces' frequency. 

The problem of discovering stochastic models through optimization has been the subject of a few recent research works. The main difficulty in this respect is that finding the optimal parameters for the stochastic process requires computing the probability of the traces issued by the model, which is algorithmically non-trivial due to the size and the likely infiniteness of the model language.

In \cite{10680664},  authors introduce an approach focused on optimizing the earth mover's stochastic conformance score of a discovered stochastic Petri net through subgradient ascent. Their approach consists of two main steps: first, they derive the stochastic language of the net by analyzing its structure to assess conformance with the event log. In the second step, they perform subgradient optimization on the loss function with respect to the model's stochastic language. This subgradient is then propagated to update the weights in the Petri net, progressively improving its alignment with the log's observed behavior. The approach has been implemented by the WAWE tool~\cite{wasserstein-spn-weight-estimation}.

In~\cite{10.1007/978-3-031-61057-8_3},  authors propose an optimization scheme that requires extracting the analytical expression for the probability of each trace issued by the model. These  expressions are obtained as solutions of $n$ ($n$ being the traces in the log) absorbing state probability problems for $n$ different discrete-time Markov chains (given by the cross-product of the stochastic reachability graph underlying the considered stochastic Petri net with a deterministic finite automaton corresponding to a trace of the log). Although effective, this approach faces scalability issues: the size of the analytical expressions extracted from the model increases dramatically with the length of the traces, making the approach unfeasible even for relatively simple models.
The approach has been implemented by the SLPN miner tool~\cite{slpn-miner}.

\section{Preliminaries}
\label{sec:prelim}

\noindent\emph{Alphabets, traces,  languages, stochastic language.}
We let  $\Sigma$ denote the alphabet of an event log's activities (we use letters to denote activities of an alphabet, e.g., $\Sigma=\{a,b,c\}$) and  $\Sigma^*$  the set of 
traces (words) composed of activities in $\Sigma$ where $\varepsilon\in\Sigma^*$ represents the empty trace. We let $Tr\subset \Sigma^*$  denote a generic set of traces built on  alphabet $\Sigma$ and $t\in Tr\subset\Sigma^*$ a trace in $Tr$, for example, $t=\langle a,b,b,a,c\rangle\in Tr\subset\{a,b,c\}^*$. A stochastic language over an alphabet $\Sigma$ is a function $L:\Sigma^*\to[0,1]$ that provides the probability of the traces such that $\sum_{t\in\Sigma^*} L(t) = 1$. 

\noindent\emph{Event log.} An event log $E$ is a multi-set of traces built on an alphabet that we denote $\Sigma_E$, i.e., $E\in Bag(\Sigma_E^*)$. Given a trace $t\in E$ we denote by $f(E,t)$ its multiplicity (i.e., its frequency in $E$). The stochastic language induced by an event log $E$ is straightforwardly obtained by computing the probability of the traces as $p(E,t)=f(E,t)/\sum_{t\in Supp(E)} f(E,t)$ where $Supp(E)$ denotes the set of unique traces in $E$, i.e., its support. The stochastic language of an event log $E$ will be denoted by $L_E$.
For $E$ an event log over an $n$-letters alphabet $\Sigma_E=\{a_1,\ldots ,a_n\}$ we denote $l_i$ ($1\leq i\leq n$) the maximum number of occurrence of letter $a_i$ in any trace of $E$ and $c_m$ the maximal length of any trace in $\Sigma_E$.

\noindent\emph{Petri net.} A labeled Petri net model is a tuple $N=(P,T,F,\Sigma,\lambda,M_0)$, where $P$ is the set of places, $T$ the set of transitions, $F:(P\times T)\cup(T\times P)\to\mathbb{N}$ gives the arcs' multiplicity (0 meaning absence of arc), $\lambda: T\to (\Sigma\cup\tau)$ associates each transition with an action ($\tau$ being the silent action) and $M_0:P\to\mathbb{N}$ is the initial marking. For $N$ a PN, we denote $RG(N)=(S,A)$, where $S=RS(N)$ is the reachability set (the set of markings reachable from the initial one, and $A\subseteq RS(N)\times RS(N)\times T$  is the set of arcs whose elements $(M,M',t)\in A$ are such that $M[t\rangle M'$, i.e. $M'$ is reached from $M$ by firing of $t$.

\noindent\emph{Workflow net.} A workflow net is a 1-safe\footnote{In any marking each place may contain at most one token} Petri net with the following structural constraints: 1) there exists a unique place, denoted \emph{source} with no incoming transition and a unique place denoted \emph{sink} with no outgoing transitions; 2) the initial marking is $M_0(source)=1$ and $M(p)=0$ for any place $p$ different from $source$ and 3) the net graph can be turned into a strongly connected one by adding a single transition outgoing place $sink$ and ingoing place $source$.

\noindent\emph{Stochastic workflow net.} In the remainder, we consider the stochastic, untimed extension of workflow nets, which we refer to as stochastic workflow nets (sWN). In practice a sWN $N_s=(P,T,F,W,\Pi,\Sigma,\lambda,M_0)$  is a generalized stochastic Petri net (GSPN)~\cite{10.1145/288197.581193} consisting uniquely of \emph{immediate} transitions, i.e., $T=T_i\cup T_t$, with $T_t=\emptyset$ ($T_t$, resp. $T_i$, being the subset of timed transitions, resp. immediate transitions), each of which is associated with a non-negative weight $W:T\to \mathbb{R}_{>0}$ while priorities are all equal ($\Pi(t)=1, \forall t\in T$), therefore in the remainder we omit $\Pi$ from the characterization of a sWN. 
In a marking $M$ transition $t$ of a sWN fires with a  probability that is a function of their weights, i.e., $P(t|M)=W(t)/\sum_{t'\in en(M)} W(t')$. The probability of firing a transition induces a probability over sequences of transition firings. Therefore, if marking $M'$ is reachable from $M$ through the sequence $M\!=\!M_1[t_1\rangle M_2[t_2\rangle~\ldots~[t_{n}\rangle M_{n+1}\!=\! M',$ then the probability of the sequence of transitions $\langle t_1,\ldots ,t_n\rangle$ (hence of the corresponding trace $\langle\lambda(t_1),\lambda(t_2),\dots,\lambda(t_n)\rangle$) starting from marking $M=M_1$ is $P(\langle t_1,\ldots ,t_n\rangle|M)=\prod_{i=1}^{n} P(t_i|M_i)$. 
We denote by $L_{N_s}$ 
the stochastic language associated with the sWN $N_s$. As in the remainder, we introduce an approach for inferring the transition weight of a sWN; we will use  $N_s(\overline{W})$ to highlight the weights parameters of sWN $N_s$, where $\overline{W}$  denote an array of transitions weights parameters.

\begin{example}
Figure~\ref{fig:ex1} shows an example of the event log and corresponding stochastic workflow net together with its stochastic language.
\vspace{-0.5cm}
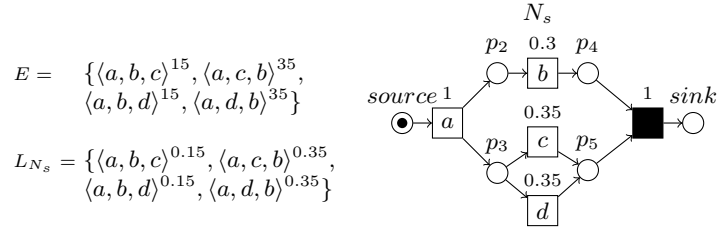
\begin{figure}[htbp]
    \begin{center}
        \begin{tabular}{cc}
            \begin{tabular}{ll}
            \scriptsize
                $E = $ & $\{ \langle a,b,c \rangle^{15}, \langle a,c,b \rangle^{35},$ \\
                & $\langle a,b,d \rangle^{15}, \langle a,d,b \rangle^{35} \}$ \\ \\
                \scriptsize
                $L_{N_s} = $ & $\{ \langle a,b,c\rangle^{0.15}, \langle a,c,b\rangle^{0.35},$ \\ 
                & $\langle a,b,d\rangle^{0.15}, \langle a,d,b\rangle^{0.35} \}$
            \end{tabular}
            &
            \begin{tabular}{c}
                $N_s$ \\
                \tikzstyle{place}=[circle,draw,minimum
size=4mm, rounded corners=0pt,scale=0.7, 
    every node/.style={transform shape}]

\begin{tikzpicture}
\node[place,tokens=1,label=above:{\footnotesize $source$}]        (source) {};
\node[node distance = 0.25 cm, transition, right=of source,label=above:\scriptsize{$1$}] (t1) {$a$};
\node[node distance = 0.5 cm, place,tokens=0, above right=of t1,label=above:$p_2$]        (p2) {};
\node[node distance = 0.5 cm,place,tokens=0, below right=of t1,label=above:$p_3$]        (p3) {};
\node[node distance = 0.25 cm, transition, right=of p2,label=above:\scriptsize{$0.3$}] (t2) {$b$};
\node[node distance = 0.25 cm,place,tokens=0, right=of t2,label=above:$p_4$]        (p4) {};    
\node[node distance = 0.5 cm, transition, below=of t2,label=above:\scriptsize{$0.35$}] (t4) {$c$};
\node[node distance = 0.5 cm, transition, below =of t4,label=above:\scriptsize{$0.35$}] (t5) {$d$};
\node[node distance = 1 cm,place,tokens=0, below=of p4,label=above:$p_5$]        (p5) {};    
\node[node distance = 2.25 cm, fill=black,transition, right =of t1,label=above:\scriptsize{$1$}] (t6) {};
\node[node distance = 0.25 cm,place,tokens=0, right=of t6,label=above:{\footnotesize $sink$}]        (sink) {};    
\draw (source) [->] to (t1);
\draw (t1) [->] to (p2);
\draw (t1) [->] to (p3);
\draw (p2) [->] to (t2);
\draw (t2) [->] to (p4);
\draw (p3) [->] to (t4);
\draw (p3) [->] to (t5);
\draw (t4) [->] to (p5);
\draw (t5) [->] to (p5);
\draw (p4) [->] to (t6);
\draw (p5) [->] to (t6);
\draw (t6) [->] to (sink);
\end{tikzpicture}\\         
            \end{tabular}
        \end{tabular}
        \vspace{-0.8cm}
        \caption{An log $E$ with its corresponding sWN ($N_s$): transition weights are depicted above each transition, while silent transitions are depicted as black rectangles. Notice that the stochastic language  $L_{N_s}$ of the sWN conforms that of the log $E$.}
        \label{fig:ex1}
    \end{center}
\end{figure}
\end{example}
\vspace{-1.1cm}
\noindent\emph{Earth Mover's Distance (EMD).} We refer to the EMD~\cite{LEEMANS2021101724} as a means to assess resemblance between the stochastic languages of logs and (sWN) models. With EMD  the cost of transforming the distribution of the log's traces into the distribution of traces issued by the model language, depends on the distances between traces which, in turns, we determine by means of the Levenshtein distance~\footnote{which measures the distance between two traces as the minimum \emph{alignment}~\cite{https://doi.org/10.1002/widm.1045}, i.e., the minimum number of single-character edits (insertion, deletion or substitution of an action) needed to change one trace into the other.}.  In order to deal with possible infiniteness of  the model's language in the remainder we to a so-called restricted  EMD (rEMD), which results by applying EMD to compare the log's traces with the subset of the model's traces consisting only of traces that belong to the log. 

\noindent\emph{HASL model checking.} 
In Section~\ref{sec:method} we introduce a formal approach to approximate the stochastic language of a sWN which is based on the HASL statistical model checking approach (HASL-SMC)~\cite{BALLARINI201553}.  
HASL-SMC  (Figure~\ref{fig:hasl}) allows for assessing sophisticated performance indicator of a (timed or untimed) GSPN model $N_s$, via  a property $\varphi\equiv({\cal A},Z)$  formally encoded by a combination of  linear hybrid automaton ${\cal A}$ and a target expression $Z$. The functioning of the framework can be summarised as follows:  a sufficiently large number of (finite) traces  are sampled (via simulation) from $N_s$ and synchronised  (\emph{on-the-fly}) with  ${\cal A}$ and those that meet the acceptance condition(s) of ${\cal A}$ are used (together with the statistics collected in the variables of ${\cal A}$)  to build  an $\epsilon$\%-confidence level estimate (with confidence interval width $\delta$)  of the  measure of interest $Z$. In the reminder we use $\mathit{CI}(N_s, \varphi,\epsilon, \delta)$ to denote such confidence interval. 
\begin{figure}[h]
    \centering
    \includegraphics[scale=0.25]{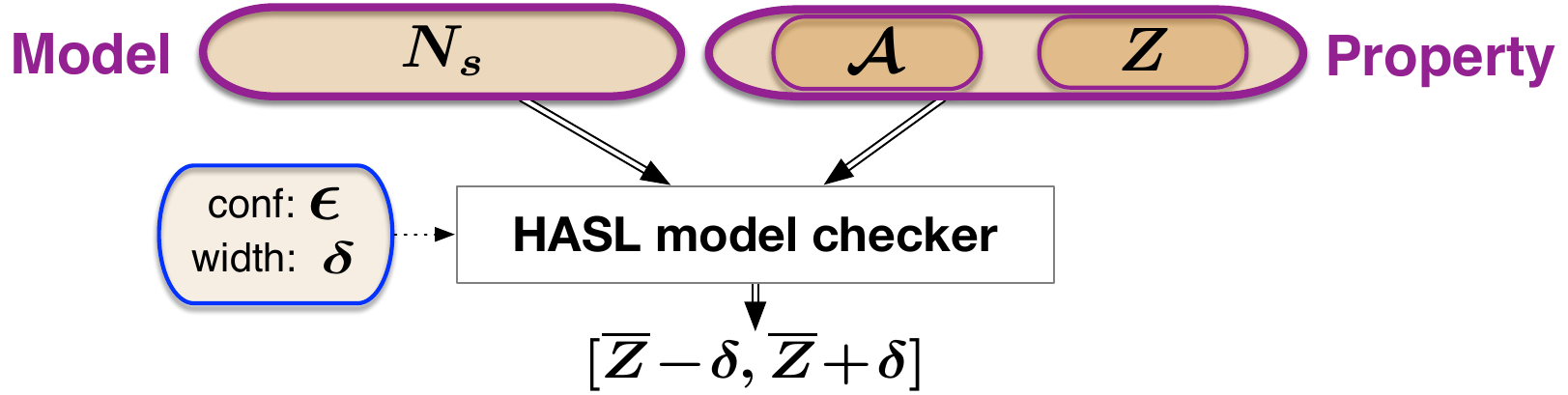}
    \caption{The HASL statistical model checking scheme}
    \label{fig:hasl}
\end{figure}

Within HASL a lynear hybrid aoutama (LHA) associated to a GSPN\footnote{whose timed transitions may be associated with non-exponential distributions.} $N_s=(P,T,F,W,\Pi,\Sigma,\lambda,M_0)$ is a tuple  $\mathcal A_{N_s}\!=\!\langle Ev, L,  I, F,  X, \flow, \Lambda, \rightarrow \rangle$, where $Ev=\Sigma\cup\{\tau\}$ is the alphabet of observed events, $L$ is a finite set of locations, with $I\subset L$, and $F\subset L$, the initial, respectively, the final (accepting) locations, $X=\{x_1,\ldots x_n\}$ a finite set of real-valued variables, $\flow:L\to (RS(N_s)\to \mathbb{R}^n)$ specifies (for each location) the rate (i.e. first derivative) with which each variable $x_i$ evolves  depending on the current marking of $N_s$, $\Lambda:L\to(RS(N_s)\to\mathbb{B})$ are the location \emph{invariants} (i.e. boolean evaluated propositions built based depending on the current marking of $N_s$) and $\rightarrow$ is a set of transitions $l\xrightarrow{Ev',\gamma, U}l'$ where $\gamma$ is an enabling guard (an inequality built on top of variables $X$), $Ev'$ is either a set of events names (i.e. the transition is \emph{synchronously} traversed on the occurrence of any reaction in $Ev'$ occurring in the path being sampled) or $\sharp$ (i.e. the transition is \emph{autonomously} traversed without synchronization) and $U$ are the variable updates.
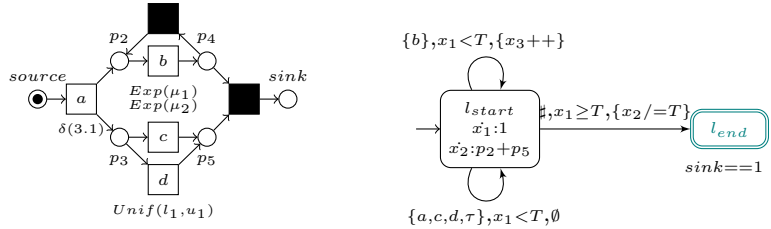
\begin{figure}[b]
    \centering
    \begin{tabular}{lr}
        \begin{tabular}{c}
            \tikzstyle{place}=[circle,draw,minimum
size=4mm, rounded corners=0pt,scale=.6]

\begin{tikzpicture}[scale=.8,minimum width=1cm]
\renewcommand{\arraystretch}{.7}
   \everymath{\scriptstyle}
\node[place,tokens=1,label=above:{\footnotesize $source$}]        (source) {};
\node[node distance = 0.25 cm, transition, right=of source,label=below:\scriptsize{$\delta(3.1)$}] (t1) {$a$};
\node[node distance = 0.3 cm, place,tokens=0, above right=of t1,label=above:$p_2$]        (p2) {};
\node[node distance = 0.3 cm,place,tokens=0, below right=of t1,label=below:$p_3$]        (p3) {};
\node[node distance = 0.25 cm, transition, right=of p2,label=below:\scriptsize{$Exp(\mu_1)$}] (t2) {$b$};
\node[node distance = 0.25 cm,place,tokens=0, right=of t2,label=above:$p_4$]        (p4) {};    
\node[node distance = 0.25 cm, transition, right=of p3,label=above:\scriptsize{$Exp(\mu_2)$}] (t4) {$c$};
\node[node distance = 0.15 cm, transition, below =of t4,label=below:\scriptsize{$Unif(l_1,u_1)$}] (t5) {$d$};
\node[node distance = 0.25 cm,place,tokens=0, right=of t4,label=below:$p_5$]        (p5) {};    
\node[node distance = 1.75 cm, fill=black,transition, right =of t1] (t6) {};
\node[node distance = .15 cm, fill=black,transition, above =of t2] (t7) {};
\node[node distance = 0.25 cm,place,tokens=0, right=of t6,label=above:{\footnotesize $sink$}]        (sink) {};    
\draw (source) [->] to (t1);
\draw (t1) [->] to (p2);
\draw (t1) [->] to (p3);
\draw (p2) [->] to (t2);
\draw (t2) [->] to (p4);
\draw (p3) [->] to (t4);
\draw (p3) [->] to (t5);
\draw (t4) [->] to (p5);
\draw (t5) [->] to (p5);
\draw (p4) [->] to (t6);
\draw (p5) [->] to (t6);
\draw (t6) [->] to (sink);
\draw (p4) [->] to (t7);
\draw (t7) [->] to (p2);
\end{tikzpicture}\\ 
        \end{tabular} &  
        \begin{tabular}{c}
\begin{tikzpicture}[scale=.775,minimum width=1cm]
\renewcommand{\arraystretch}{.7}
   \everymath{\scriptstyle}
   
\draw (-3,1.5) node[initial,left, initial text=,draw,rounded corners] (l0) { \begin{tabular}{@{}c@{}}{$l_{start}$}\\  $\dot{x_1}:1$ \\ $\dot{x_2}:p_2+p_5$\end{tabular} };
\draw (0.25,1.5) node[draw,rounded corners,accepting,color=teal] (l1) { \begin{tabular}{@{}c@{}}{$l_{end}$}\end{tabular} }; 

\node (acc) [text width =1.2cm,below =.05cm of l1] {$sink==1$};

\draw [-latex'] (l0) -- (l1) node [midway, above,sloped] {$\sharp$,$x_1\geq T$,$\{x_2/=T\}$ };
\draw [-latex'] (l0) .. controls +(115:15mm) and +(75:15mm) .. (l0) node [midway ,above] { $\{b\}$,$x_1<T$,$\{x_3++\}$ };
\draw [-latex'] (l0) .. controls +(-115:15mm) and +(-75:15mm) .. (l0) node [midway ,below] { $\{a,c,d,\tau\}$,$x_1< T$,$\emptyset$ };




{}; 
 
\end{tikzpicture}
        \end{tabular} 
    \end{tabular}    
    \caption{An LHA synchronisying with a (timed) GSPN}
    \label{fig:lhaexample}
\end{figure}

\begin{example} \label{ex:lha}
Figure~\ref{fig:lhaexample} shows an example of  GSPN model $N_s$ (left) and a corresponding synchronising LHA ${\mathcal A_{N_s}}$ (right). Model $N_s$ is a timed extension of that in Figure~\ref{fig:ex1} (right) such that transitions labeled with actions \texttt{a}, \texttt{b}, \texttt{c} and \texttt{d} are timed and associated to different kind of delay distribution (Dirac, Exponential and Uniform) while silent transitions are assumed to be immediate. The LHA has an initial  ($l_{start}$) and a final ($l_{end}$) location and uses three variables $X=\{x_1,x_2,x_3\}$ whose flows (rates) are $\flow(l_{start})\!=\!(1,p_2+p5,0)$ and $\flow(l_{end})\!=\!(0,0,0)$ which means  
that  $x_1$ is used as clock (rate $\dot{x}_1\!=\!1$),  $x_2$ is used to hold the integral of the sum of tokens in places $p_2$ and $p_5$ ($\dot{x}_2\!=\!p_2+p_5$) while $x_3$ is not evolving as ${\mathcal A_{N_s}}$ spends time in $l_{start}$. 
Finally ${\mathcal A_{N_s}}$ has two self-loop \emph{synchronised} transitions $l_{start}\xrightarrow{\{b\},x_1<T,x_3+\!=\!1}l_{start}$, $l_{start}\xrightarrow{\{a,c,d,\tau\},x_1<T,\emptyset}l_{start}$, traversed by synchronisation with occurrences of transitions of model $N_s$
and an \emph{autonomous} transition $l_{start}\xrightarrow{\sharp,x_1\!=\!T,x_2/=T}l_{end}$, traversed autonomously as soon as its guard ($x_1\!\geq\! T$) is satisfied, on condition that the invariant $\Lambda(l_{end}):sink\!==\!1$ of destination location $l_{end}$ is satisfied. 
The traces accepted by ${\cal A}_{N_s}$ are then used to assess relevant performance measure formally given by an expression $Z$, such as, for example, $Z=\mathit{PDF}(last(x_2),0.01,0,2)$ that represents the PDF of the \emph{average  number of tokens contained in places $p_2$ and $p_5$ within time at least $T$} approximated using $[0,2]$ as support set and discretizing $[0,2]$ with buckets. 
We give a detailed description of the synchronisation between model $N_s$ and automaton ${\mathcal A}_{N_s}$ and of the grammar of target expression $Z$  in Section~\ref{sec:hasl}.  Notice that differently from the example in Figure~\ref{fig:lhaexample} within this paper HASL is referred exclusively to the sWN subclass of GSPN (i.e. untimed models) and to $PDF$ target expressions (i.e. expression to approximate a probability distribution function).  
\end{example}

\noindent\emph{Approximate Bayesian Computation.} The optimization framework we introduce is obtained as an adaptation of  Approximate Bayesian Computation (ABC)~\cite{Marin2011,Sisson2018}, i.e., a family of   methods that taking from a prior distribution $\pi(.)$ over the  model's parameter space ($\overline{W}$  in the sWN case), allows one to obtain an estimate of the posterior distribution  denoted $\pi_{ABC,\epsilon}(.)$, whose precision depends on  the chosen  tolerance  $\epsilon$ value. 
To overcome slow convergence issues of the simple \emph{rejection sampling} ABC algorithm we considered  its sequential Monte Carlo extension, named ABC-SMC~\cite{Beaumont2008}. We give a detailed description of the ABC method  in Appendix~\ref{app:ABC}. 

\section{Method}
\label{sec:method}
We introduce a simulation-based method for  discovery of optimal weight parameters of a sWN model $N_s$. The method combines the HASL-SMC procedure for approximating the stochastic language $L_{N_s}$ issued by  $N_s$, with an  adaptation of the ABC-SMC scheme to  identify the ``best'' transitions weights  for $N_s$. 

\subsection{HASL-based approximation of a sWN stochastic language}
\label{sec:haslsle}
We present an HASL formula  (Definition~\ref{def:haslsle}), that relying on a dedicated \emph{stochastic language detector} (hybrid) automaton ${\cal A}_{\mathit{sld}(E)}$ (Definition~\ref{def:haslsld})   allows one to obtain an arbitrarily precise approximation of  the stochastic language $L_{N_s}$. 



To this aim we first define a mapping through which words over an alphabet $\Sigma$ are mapped to unique integer values. 

\begin{definition}[Word mapping]
\label{def:wordmapping}
Given an alphabet $\Sigma$ consisting of $n$ letters ($|\Sigma|=n$) and an injective function $f_l:\Sigma\to\mathbb{N}$ we define the word mapping function $w_m:\Sigma^*\to \mathbb{N}$ as: 
\begin{equation}
\label{eq:wordmapping} 
w_m(\sigma)=\sum_{i=1}^{|\sigma|}f_l(\sigma[i])\cdot n^{i-1}
\end{equation}

\end{definition}
Notice that, trivially, mapping $w_m$ is injective as, by hypothesis, the letters mapping $f_l$ is injective. 

\begin{example}
Let $\Sigma=\{a,b,c\}$ be an alphabet, and let us assume $f_l(a)=1$, $f_l(b)=2$ and $f_l(c)=3$ as letters' mapping, then the following are examples of mapping of words in $\Sigma^*$: $w_m(aa)=1\cdot 3^0 + 1\cdot 3^1=4$, $w_m(cb)=3\cdot 3^0 + 2\cdot 3^1=9$, $w_m(abc)=1\cdot 3^0 + 2\cdot 3^1+ 3\cdot 3^2=34$, $w_m(cba)=3\cdot 3^0 + 2\cdot 3^1+ 1\cdot 3^2=18$.

\end{example}

\vskip -1ex
In the remainder given a finite event log  $E\in \mathit{Bag}(\Sigma_E^*)$ and a word mapping $w_m$ (defined on alphabet $\Sigma_E$), we denote $w_m(E)\subset\mathbb{N}$ the set of naturals to which the words of $Supp(E)$ are mapped.  

\paragraph{Convex remapping.}
As mapping~(\ref{eq:wordmapping}) commonly yields a non-convex (sparse) support set with large\footnote{Exponential in the size of the word.}  supremum, which would negatively impact the HASL-based  confidence interval estimation of the corresponding probability density function (PDF) expression\footnote{In this context, what we deal with are probability mass functions (PMF). We still use PDF because it is a keyword in \cosmos~covering both PDFs and PMFs.},
we use a convex re-mapping scheme through which each word of a finite log $E$ is mapped back over the convex interval $\{0,1,2,...,|Supp(E)|\!-\!1\}$. To this aim we first compute (offline) $w_m(E)$ and then re-map each element in $w_m(E)$ to a corresponding value in $\{0,1,2,...,|Supp(E)|\!-\!1\}$. Therefore in the remainder for a word $\sigma\in Supp(E)$ we denote its convex remapping $w_{cm}(w_m(\sigma))\in\{0,1,2,...,|Supp(E)|\!-\!1\}$.


\begin{definition}[Stochastic language detector automaton] 
\label{def:haslsld}
Given $N_s$, a sWN discovered from  an event log  $E$ with  alphabet $\Sigma_E=\{a_1,\ldots ,a_n\}$ and given an injective  mapping $f_l:\Sigma_E\to\mathbb{N}$, the \emph{stochastic language estimator} LHA 
 $\mathcal {\cal A}_{sld(E)}\!=\!\langle Ev, L,  I, F,  X, \flow, \Lambda, \rightarrow \rangle$ is defined as follows: event set $Ev=\Sigma_E\cup\{\tau\}$, locations $L=I\cup F$ with $I=\{l_{start}\}$ and $F\!=\!\{l_{end}\}$, variables set $X\!=\!\{w,w_{c},c,c_1,\dots,c_n \}$ consisting of the following $n+3$ (integer) variables, 
  \begin{itemize}
     \item $w$: mapping of detected word
     \item $w_{c}$: convex mapping of detected word
     \item $c$: length of detected word
     \item $c_i$ ($1\!\leq\! i \!\leq\! n)$: number of occurrences of letter $a_i$  in the detected word
 \end{itemize}

\noindent
each with constant rate of evolution in every location $\flow(l)=(0,\ldots,0)$ (~$\forall l\in L$), $\Lambda(l_{start})=\texttt{true}$, location invariants $\Lambda(l_{end})=(sink\!==\!1)$ and transition set $\rightarrow$ consisting the following $n+1$ transitions: 
 \begin{itemize}
    \item $n$ self-loop \emph{synchronised} transitions (with $1\leq i\leq n$) defined  as follows: 
\[
l_{start}\xrightarrow{\{a_i\}, c_i\!\leq l_i\land (\sum_i c_i)\leq c_m, \{w+=f_l(a_i)\cdot n^c; c+=1;c_i+=1;\}}l_{start}
\]
where $n$, $l_i$ and $c_m$ are constants referred to the log $E$ ($n$ number of letters, $l_i$ maximum number of the $i$-th letter in any word of $E$, $c_m$ length of the longest word in $E$).
\item $|Supp(E)|$ \emph{autonomous} transitions (with $1\leq i\!\leq\! |Supp(E)|$)  defined as 
$$l_{start}\xrightarrow{\sharp, w == w_m(\sigma_i), \{w_c=w_{cm}(w_m(\sigma_i))\}}l_{end}$$ 
where $\sigma_i\in Supp(E)$ is a unique trace of $E$, $w_m(\sigma_i)\in \mathbb{N}$ is its mapping and $w_{cm}(w_m(\sigma_i))\in \mathbb{N}$ its convex re-mapping.

\end{itemize}


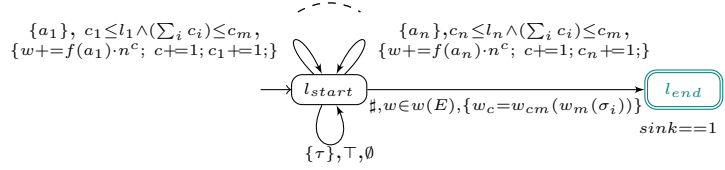
\begin{figure}
    \centering
\begin{tikzpicture}[scale=.99,minimum width=1cm]
\renewcommand{\arraystretch}{.7}
   \everymath{\scriptstyle}
   
\draw (-4,1.5) node[initial,left, initial text=,draw,rounded corners] (l0)  { {$l_{start}$} };
\draw (0.25,1.5) node[draw,rounded corners,accepting,color=teal] (l1) { \begin{tabular}{@{}c@{}}{$l_{end}$}\end{tabular} }; 



\draw [-latex'] (l0) -- (l1) node [midway, below,sloped] {$\sharp,w\in w(E),\{w_c=w_{cm}(w_m(\sigma_i))\}$ };


\draw [-latex'] (l0) .. controls +(135:10mm) and +(115:10mm) .. (l0) node (arc1) [pos=0.6] {\hskip -30ex   \begin{tabular}{@{}c@{}}$\{{a_1}\}$, $c_1\leq l_1\land (\sum_i c_i)\leq c_m$,\\ 
$\{w += f(a_1)\cdot n^c;$  $c +\!= 1; \, c_1 +\!= 1; \}$ \end{tabular}};

\draw [-latex'] (l0) .. controls +(45:10mm) and +(65:10mm) .. (l0) node (arcn) [pos=0.6] {\hskip 30ex \begin{tabular}{@{}c@{}} $\{{a_n}\}$,$c_n\leq l_n\land (\sum_i c_i)\leq c_m$, \\$\{w+=f(a_n)\cdot n^c;$  $c +\!= 1; \, c_n +\!= 1; \}$ \end{tabular}};

\draw [-latex'] (l0) .. controls +(245:10mm) and +(295:10mm) .. (l0) node (arcsilent) [below,pos=0.7] {$\{{\tau}\}$,$\top$,$\emptyset$ };

\draw [line width=.5pt,dashed](arc1.north) to [bend left=30] (arcn.north);

\node (acc) [text width =1.2cm,below =.05cm of l1] {$sink==1$}



{}; 
 
\end{tikzpicture} 
    \caption{The stochastic language detector automaton ${\cal A}_{sld(E)}$ 
    }
    \label{fig:hasl_sle}
\end{figure}
\end{definition}

\paragraph{Description.} 
The goal of automaton ${\cal A}_{sld(E)}$ (depicted in Figure~\ref{fig:hasl_sle}\footnote{For the sake of space we slightly abuse the LHA syntax and  subsume the $|Supp(E)|$ transitions $l_{start}\rightarrow l_{end}$ described in Definition~\ref{def:haslsld} by a single, semantically equivalent,  transition   with guard $w\in w_m(E)$.}) is to detect, among the traces issued by  model $N_s$  it synchronies with, those that belong to the log $E$ by retrieving their corresponding mapping. 
To this aim ${\cal A}_{sld(E)}$  is equipped with  $n\!+\!3$ integer variables whose goal is 1) to store (variable $w$) the mapping (as per  Definition~\ref{def:wordmapping})  of the trace currently being scanned, 2) to count the occurrences of each activity $a_i$ (variables $c_i$, $1\!\leq\! i\!\leq\! n$) as it is observed on the scanned trace, 3) to store the length of the word (in terms of total number of observed actions $a_i$ (variable $c$) and, finally, 4) to store the convex remapping (variable $w_c$)  when the scanned trace yield a word of $E$. 
Each observed activity is detected by traversal of the corresponding synchronized self-loop arc on the initial location ($l_{start}$). Notice that on traversal of the self-loop arc corresponding to activity $a_i$, the value of $w$ is added up with  $f_l(a_i)\cdot n^c$ while the $a_i$ activity counter $c_i$ is incremented. 

Scanning of the traces issued by a sWN model is guaranteed to terminate by either accepting or rejecting the currently observed trace. A trace $\sigma\in\Sigma_E^*$ is accepted if and only if 1) it is generated by a sequence of transitions $\langle t_1,\ldots ,t_m\rangle$ whose last transition $t_m$ reaches the deadlock marking by adding a token in place \emph{sink} (i.e., invariant  $sink\!==\!1$ of final location $l_{end}$) and 2) it belongs to the event log $\sigma\in supp(E)$ (that is, if its mapping $w$ is in $w_m(E)$, which is captured by  the guard $w\in w_m(E)$ on $l_{start}\rightarrow l_{end}$). Any other trace $\sigma\in\Sigma_E^*\setminus{supp(E)}$ results in the automaton to block and, hence, it is rejected (thus its mapping value $w$ is discarded). Notice that in order to rule out infinite traces (that may result from a net that contains loops) and more generally to shorten the  trace detection process, each self-loop, synchronous, $a_i$ arc is enabled only on condition that 1) the number of $a_i$ observed on the current trace is not above the maximum number $l_i$ computed for the traces in $E$ (captured by constraint $  c_i\leq l_i$) and 2) the length of the current trace does not trespasses $c_m$, that is, the length of the longest trace in $E$ (captured by constraint $\sum_i c_i\leq c_m$). 

By definition, automaton ${\cal A}_{sld(E)}$ enjoys the property that the language $L_{N_s\!\times\! {\cal A}_{sld(E)}}$, issued by the product process $N_s\times {\cal A}_{sld(E)}$, 
is contained in the event log $E$ from which the model $N_s$ has been discovered.  We state this property in Theorem~\ref{theo:detection}.

\begin{theorem}
\label{theo:detection}
    Let $N_s$ be a sWN discovered with a discovery algorithm $\mathit{Alg}$ from an event log $E\in \mathit{Bag}(\Sigma_E^*)$ then $L_{N_s\!\times\! {\cal A}_{sld(E)}}\subseteq supp(E)$.  
    \proof Straightforward consequence of HASL operational semantics. 
\end{theorem}

\begin{lemma}
    If  algorithm $\mathit{Alg}$ used to discover $N_s$ has fitness   1 then $L_{N_s\!\times\! {\cal A}_{sld(E)}}= supp(E)$.  
\end{lemma}

As a consequence of Theorem~\ref{theo:detection} by combining automaton ${\cal A}_{sld(E)}$ with target expression $\mathit{PDF}(last(w),s,l,h)$ (with adequate buckets and support set  parameters $s$, $l$ and $h$) we obtain a confidence interval estimator of the probability distribution with which model $N_s$ generates the traces of  log $E$.


\begin{definition}[HASL stochastic language estimator formula] 
\label{def:haslsle}
    Given $N_s$ a sWN discovered from  an event log  $E$ we define the HASL  \emph{stochastic language estimator}  formula $\varphi_{sle(E)}\equiv(\mathcal {\cal A}_{sld(E)},\mathit{PDF}(last(w_c),1,0,|Supp(E)|\!-\!1))$, where $\mathcal {\cal A}_{sld(E)}$ is the stochastic language detector automaton (Definition~\ref{def:haslsld}), $w$ its corresponding (word mapping)  variable. Notice that since $w_c$ is a discrete random variable  in the target expression $\mathit{PDF}(last(w_c),1,0,|Supp(E)|\!-\!1))$ we use buckets of size $s\!=\!1$, while we use $[l,h]\!=\![0,|Supp(E)|\!-\!1]$ as support set for the sought approximation of the PDF of $w_c$.
\end{definition}
\subsection{An ABC framework for optimised stochastic process discovery}
\label{ABC4stochasticconformance}

To discover  optimal weights $\overline{W}$ for a sWN model $N_s$ 
(mined from an event log $E$ with alphabet $\Sigma_E$), 
we introduce an adaptation of the ABC sequential Monte Carlo (ABC-SMC) parameter inference approach~\cite{Beaumont2008} in which we plug in the HASL-based stochastic language approximation method (Section~\ref{sec:haslsle}) as a mean to retrieve the stochastic language issued by the model's instance  $N_s(\overline{W})$ that corresponds to a specific parameter vector $\overline{W}$. 
In our context, the parameter space is $k$-dimensional ($k=|T|$ being the number of transitions of $N_s$) and its elements are  vectors $\overline{W}\in [0,1]^{k}$ corresponding to the weights of the transitions. The optimization objective is to minimize the rEMD distance between the stochastic language $L_{N_s(\overline{W})}$ issued by model $N_s(\overline{W})$ and the stochastic language $L_E$ of the log $E$. 
Following the ABC-SMC scheme, the novel algorithm we introduce (Algorithm~\ref{alg:abc_pmc}) operates over a sequence of $m\in\mathbb{N}$ telescopic layers $i$ ($1\leq i\leq m$) with decreasing tolerances $\epsilon_i$, such that $\epsilon_1>\epsilon_2>\ldots >\epsilon_m$\footnote{See Appendix~\ref{appendix:abcsbc}}.
 
\begin{algorithm}[h!]
\scriptsize
\caption{ABC-SMC for stochastic process discovery}\label{alg:abc_pmc}
\begin{algorithmic}
    \Require{$E$ (event log), $\Sigma_E$ (alphabet), $f_l(\Sigma_E)$ (activity mapping), $N_s(\overline{W})$ (sWN mined from $E$ with weights $\overline{W}$), $n$ (particles), $\epsilon_1$ (first layer threshold), $\zeta$ (improvement threshold), $K(\cdot|\cdot)$ (transition Kernel), $\epsilon$ (confidence-level), $\delta$ (interval-width)}
    \Ensure{$(\overline{w}_j)_{1 \leq j \leq n}$ drawn from $\pi_{ABC,\epsilon_M}$}
    \State $\triangleright$ {We use $L_{N_s(\overline{w})} = \mathit{CI}( N_s(\overline{w}),\varphi{sle(E)},\epsilon,\delta)$ to denote the stochastic language of $N_s(\overline{w})$ computed by simulation.}
    \State Iteration $i = 1$: Find $(\overline{w}_j^{(1)})_{1 \leq j \leq n}$ using the sWN adaptation of ABC rejection sampling (Algorithm~\ref{alg:abc}), which employs the threshold $\epsilon_1$ as the particle acceptance parameter.    
    \State $(\delta_j)_{1 \leq j \leq n} \gets \frac{1}{n}$
    \Repeat
        \State $i \gets i+1$
        \State $\epsilon_i \gets 
        Median\left( rEMD\left(L_E,L_{N_s\left(\overline{w}_j^{(i-1)}\right)}\right)_{1 \leq j \leq n} \right)$
        
        \For{$j = 1:n$}
            \Repeat
                \State Take $\overline{w}_j^{'}$ from $(\overline{w}_j^{(i-1)})_{1 \leq j \leq n}$ with probabilities $(\delta_j)_{1 \leq j \leq n}$
                \State $\overline{w}_j^{(i)} \sim K(.|\overline{w}_j^{'})$
            \Until{$\mathit{rEMD}\left(L_E,L_{N_s\left(\overline{w}^{(i)}_j\right)}\right) < \epsilon_i$}
            \State $\delta_j \gets \pi(\overline{w}_j^{(i)} ) \times \big(\sum^{n}_{j'=1} \delta_{j'}^{(i-1)} K(\overline{w}_j^{(i)} | \overline{w}_{j'}^{(i-1)})\big)^{-1}$
        \EndFor
    \State Normalize $(\delta_{j})_{1 \leq j \leq n}$
    \Until{$\epsilon_i -
    Median\left( rEMD\left(L_E,L_{N_s\left(\overline{w}_j^{(i)}\right)}\right)_{1 \leq j \leq n} \right)< \zeta$}
\end{algorithmic}
\end{algorithm}

At each layer $i$, a user-defined number ($n$) of \emph{particles} (in our case, weight vectors $\overline{W}$) are iteratively sampled from a given probability distribution until the measured distance (in this case, the rEMD between the event log language and the language generated by the model with weights $\overline{W}$) falls within the tolerance level $\epsilon_i$. At the first layer ($i=1$), the initial $n$ particles ($\overline{w}^{(1)}_1, \ldots, \overline{w}^{(1)}_N$) are selected using a stochastic process discovery adaptation of the simple ABC rejection sampling method (Algorithm~\ref{alg:abc} given in Appendix~\ref{app:ABC}). In this step, weight vectors 
are sampled from a $k$-dimensional uniform prior distribution $\mathcal{U}(0,1)^k$, with a sufficiently permissive user-defined tolerance $\epsilon_1$ to ensure a satisfactory acceptance rate.

At each successive level ($i \geq 2$), the $n$ particles are selected through a two-step probabilistic procedure. First, a particle from the set accepted at the previous level, $(\overline{w}^{(i-1)}_j)_{1\leq j\leq n}$, is randomly selected based on its probabilistic weight, $(\delta_j)_{1\leq j\leq n}$, where $\delta_j$ represents the probability of selecting particle $j$. Next, a new particle for the current level is sampled using a kernel distribution $K(.|\overline{w}'_j)$ centered around the previously selected particle $\overline{w}'_j$. 
As the weights of a sWN must be non-negative, we employed a $k$-dimensional normal distribution $\mathcal{N}(a,b)$, truncated to $[0,1]$. The mean $a$ and covariance matrix $b$ at the $i$-th layer  are derived from the collection of $n$ particles accepted at the previous $(i\!-\!1)$-th layer~\cite{filippi2012optimality}. 
We stress that in Algorithm~\ref{alg:abc_pmc} we apply an adaptive tolerance scheme, that is: the tolerance $\epsilon_i$ for each layer $i \geq 2$, rather then being a required input  constant of the algorithm (as in the original ABC-SMC method~\cite{Beaumont2008}),  is computed as the median of the rEMD distances of the particles accepted at the previous layer $i-1$. 
This dynamic approach allows for more flexibility and a better control of the parameter search process as the required $m$ threshold levels ($\epsilon_i$) of the original ABC-SMC method are replaced by a single \emph{improvement threshold} parameter ($\zeta$) used to establish when to terminate the search process. Specifically, at the end of each layer, the algorithm determines whether to proceed to the next layer or terminate the procedure by comparing the prospective tolerance of the next layer with the current one (i.e., $\epsilon_i -
    Median( rEMD(L_E,L_{N_s(\overline{w}_j^{(i)})})_{1 \leq j \leq n} )< \zeta$). If the improvement in tolerance is not significant (specifically, if it falls below the user-defined improvement threshold $\zeta$), the algorithm has likely reached a local minimum at the current layer and terminates as continuing to the next layer would not result in significantly better particles.
Finally, notice that  such adaptive tolerance scheme  ensures  convergence, as the tolerance for the next layer is strictly lower than that of the preceding layer ($\epsilon_{i-1} > \epsilon_{i}$). This property holds because the median is calculated solely from particles whose rEMD distances are strictly less than the previous tolerance $\epsilon_{i-1}$. 


\section{Experiments and results}
\label{sec:results}

We developed a prototype implementation of the HASL-based ABC-SMC parameter inference scheme described in Section~\ref{sec:method}. To validate it, we conducted two types of experiments. The first (Section~\ref{result:hasl}) evaluates the precision and cost of the HASL-based stochastic language estimator (Definition~\ref{def:haslsle}) in isolation. The second (Section~\ref{result:smcabc}) tests the ability of ABC-SMC to infer weights such that the stochastic language of the sWN aligns with that of the log. All experiments were conducted on real-life event logs of varying complexity\footnote{From the Business Process Intelligence challenge~\url{https://data.4tu.nl/}}, with corresponding WN models discovered using the inductive miner algorithm \cite{DBLP:conf/apn/LeemansFA13}, guaranteeing fitness equal to 1, that is, models that reproduces all traces of the log.

\noindent
\emph{Prototype tool.} The tool, written in Python, uses the simulator generated by \cosmos~for a GSPN instance to compute the approximate stochastic language and integrates it with a Python-based ABC-SMC implementation for parameter inference. Source code and results are publicly available in the Git repository at \gitlink. All experiments were conducted on an Ubuntu machine with a 2.60GHz CPU.

\subsection{Accuracy of HASL-based stochastic language estimates}
\label{result:hasl}
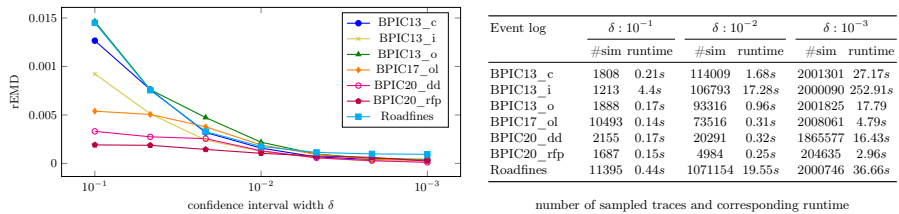
\begin{figure}[t]
    \centering
    \begin{tabular}{cc}
     \begin{tabular}{c}
     \resizebox{.5\textwidth}{!}{
        \begin{tikzpicture}
            \begin{axis}[
                width=12cm, height=6cm,
                xlabel={confidence interval width $\delta$},
                ylabel={rEMD},
                xmode=log,
                log basis x={10},
                xtick={10^-1, 10^-2, 10^-3},
                xticklabels={$10^{-1}$, $10^{-2}$, $10^{-3}$},
                ytick={0, 0.005, 0.01, 0.015},
                yticklabels={$0$, $0.005$, $0.001$, $0.015$},
                scaled y ticks=false,
                mark options={solid},
                legend pos=north east,
                x dir=reverse,
            ]
        
                \addplot[mark=*, blue, thick] coordinates {
                    (0.1, 0.012664975655563297)
                    (0.046415888336127795, 0.007624845182960445)
                    (0.021544346900318843, 0.003190184729177608)
                    (0.01, 0.0015811286664432836)
                    (0.0046415888336127815, 0.000666347229086146)
                    (0.0021544346900318847, 0.00040869474550904076)
                    (0.001, 0.0003772585627204316)
                };
                \addplot[mark=x, yellow!80!black, thick] coordinates {
                    (0.1, 0.009235690807200441)
                    (0.046415888336127795, 0.005124355708500474)
                    (0.021544346900318843, 0.002402047853550451)
                    (0.01, 0.0012411867497504447)
                    (0.0046415888336127815, 0.0005718487375504332)
                    (0.0021544346900318847, 0.00023880658745041092)
                    (0.001, 0.0002816068453004442)
                };
                \addplot[mark=triangle*, green!50!black, thick] coordinates {
                    (0.1, 0.014642564654450421)
                    (0.046415888336127795, 0.007608547632900464)
                    (0.021544346900318843, 0.004734568159100429)
                    (0.01, 0.0021882340405504504)
                    (0.0046415888336127815, 0.0009621010139504447)
                    (0.0021544346900318847, 0.0005116102633504443)
                    (0.001, 0.0004184482199005109)
                };
                \addplot[mark=diamond*, orange, thick] coordinates {
                    (0.1, 0.00539026096020005)
                    (0.046415888336127795, 0.005053397558100066)
                    (0.021544346900318843, 0.0037783847298900426)
                    (0.01, 0.00190414703634003)
                    (0.0046415888336127815, 0.0009333343036800267)
                    (0.0021544346900318847, 0.00042651090495005663)
                    (0.001, 0.00035188258986003997)
                };
                \addplot[mark=o, magenta, thick] coordinates {
                    (0.1, 0.003309638949550023)
                    (0.046415888336127795, 0.0027457694799500223)
                    (0.021544346900318843, 0.002554973599500011)
                    (0.01, 0.001303871386800039)
                    (0.0046415888336127815, 0.0005776344133500278)
                    (0.0021544346900318847, 0.00028861761505004997)
                    (0.001, 0.00010563004110001112)
                };
                \addplot[mark=pentagon*, purple, thick] coordinates {
                    (0.1, 0.0019115008754002224)
                    (0.046415888336127795, 0.001869975031300228)
                    (0.021544346900318843, 0.0014438891904502277)
                    (0.01, 0.0010451592234502333)
                    (0.0046415888336127815, 0.0007558558697002444)
                    (0.0021544346900318847, 0.0006255233380002387)
                    (0.001, 0.00024294821915024986)
                };
                \addplot[mark=square*, cyan, thick] coordinates {
                    (0.1, 0.014497603188107718)
                    (0.046415888336127795, 0.007553939456387722)
                    (0.021544346900318843, 0.0032822724896351685)
                    (0.01, 0.0017499999308327372)
                    (0.0046415888336127815, 0.0011346547310451848)
                    (0.0021544346900318847, 0.0009793325772926582)
                    (0.001, 0.00094562897658942531)
                };
                \legend{BPIC13\_c, BPIC13\_i, BPIC13\_o, BPIC17\_ol, BPIC20\_dd, BPIC20\_rfp, Roadfines}
            \end{axis}
        \end{tikzpicture}}
     \end{tabular}
     &  
     \begin{tabular}{c}
         \resizebox{.45\textwidth}{!}{
        \begin{tabular}{l p{3ex} c c p{1ex} c c p{1ex} c c}
            \toprule
            Event log & & \multicolumn{2}{c}{$\delta : 10^{-1}$} & & \multicolumn{2}{c}{$\delta : 10^{-2}$} & & \multicolumn{2}{c}{$\delta : 10^{-3}$} \\
            \cmidrule(){3-4} \cmidrule(){6-7} \cmidrule(){9-10}
             & & \#sim & runtime & & \#sim & runtime & & \#sim & runtime \\
            \midrule
            BPIC13\_c    & & $1808$  & $0.21s$ & & $114009$  & $1.68s$  & & $2001301$ & $27.17s$  \\
            BPIC13\_i    & & $1213$  & $4.4s$  & & $106793$  & $17.28s$ & & $2000090$ & $252.91s$ \\
            BPIC13\_o    & & $1888$  & $0.17s$ & & $93316$   & $0.96s$  & & $2001825$ & $17.79$ \\
            BPIC17\_ol   & & $10493$ & $0.14s$ & & $73516$   & $0.31s$  & & $2008061$ & $4.79s$ \\
            BPIC20\_dd   & & $2155$  & $0.17s$ & & $20291$   & $0.32s$  & & $1865577$ & $16.43s$ \\
            BPIC20\_rfp  & & $1687$  & $0.15s$ & & $4984$    & $0.25s$  & & $204635$  & $2.96s$ \\
            Roadfines    & & $11395$ & $0.44s$ & & $1071154$ & $19.55s$ & & $2000746$ & $36.66s$ \\
            \bottomrule \\ 
            \multicolumn{10}{c}{number of sampled traces and corresponding runtime}
        \end{tabular}}
     \end{tabular}
    \end{tabular}
    \caption{HASL-based stochastic language approximation: accuracy and runtime.}
    \label{fig:cosmos_results}
\end{figure}

To assess the accuracy of the HASL-based estimation of the stochastic language of a given sWN model, we run a number of experiments using the \cosmos~model checker. 
For each log, we discovered the corresponding WN $N$ and then obtained several instances $N_s(\overline{W}_i)$ of the corresponding sWN by randomly generating weight vectors  $\overline{W}_i\in[0,1]^{|T|}$ for the transitions $T$ of $N$. 
For each 
model instance $N_s(\overline{W}_i)$, we approximated its stochastic language, evaluating the formula $\varphi_{\mathit{sle}(E)}$ on \cosmos, using $\epsilon=99\%$ as confidence level and varying the length $\delta$ of the confidence interval (note that $\epsilon$ and $\delta$ determine the number of necessary simulation runs). The obtained approximation was compared, in terms of rEMD, to the exact stochastic language of the model, computed via the reachability graph unfolding method introduced in~\cite{DBLP:journals/corr/abs-2406-10817}. The results are shown on the left in Figure~\ref{fig:cosmos_results}. The plot indicates an increasing precision of the HASL approximations of $L_{N_s(\overline{W})}$ as the length of confidence interval $\delta$ is decreased, with the approximated and the exact language getting essentially indistinguishable (rEMD$\,\approx 0$ with $\delta= 10^{-3}$) for all logs.

The table in Figure~\ref{fig:cosmos_results} details the computational costs of the experiments, both in terms of the number of simulations and the corresponding runtime for different values of  $\delta$. Combined with the plot in Figure~\ref{fig:cosmos_results} the values allow one to quantify the trade-off between precision and computational costs. Notice that even with a relatively large interval size ($\delta\!=\! 10^{-1}$) we obtain precise estimates of the language (rEMD $<0.015$) in a few tenths of a second (except for one log).  Clearly, an increased precision is paid in terms of longer runtime which yet remains below one second for 5 out of 7  of the considered logs when an excellent precision is chosen (i.e., $\delta\!=\! 10^{-2}$ resulting in a rEMD $< 0.001$ for all logs).



\subsection{HASL based ABC-SMC stochastic process discovery}
\label{result:smcabc}

Table \ref{tab:exp1} compares the outcomes of sWN parameter estimation resulting from the ABC-SMC procedure outlined in Section~\ref{sec:method} (column ``ABC-SMC opt'') with (1) the reachability graph unfolding optimization method~\cite{DBLP:journals/corr/abs-2406-10817} (column ``Unfolding opt''), (2) multiple weight estimators derived from statistic activity relations~\cite{DBLP:conf/icpm/BurkeLW20} (column ``Weight estimators''), (3) the Wasserstein weight estimator~\cite{10680664} (column ``WAWE'') and (4) the stochastic labeled Petri net miner~\cite{10.1007/978-3-031-61057-8_3} (column ``SLPN Miner'').
Column ``$|Tr|$'' indicates the number of unique traces in the log and ``$|T|$'' the number of transitions in the corresponding mined WN (i.e., the dimensionality of the parameter space). The ``rEMD'' columns report the distance between the stochastic language of the optimized model and that of the log. 

For ``Weight estimation'', among the six estimators introduced in \cite{DBLP:conf/icpm/BurkeLW20}, only the one yielding the lowest rEMD is reported.
Experiments with the WAWE tool require setting of five hyper-parameters that may strongly affect the result quality and in this respect we resorted to  reference settings given in~\cite{10680664}. Further, following the experimental procedure  in~\cite{10680664} and due to variability in the output, we ran each WAWE experiment 30 times, computed the rEMD for each run, and reported the median rEMD. 
Except for \texttt{BPIC17\_ol}, optimization with SLPN Miner consistently failed to terminate due to the symbolic representation of trace probabilities becoming computationally infeasible in complex scenarios. In the ``ABC-SMC opt'' column, $\zeta$ indicates the improvement threshold used during parameter inference and $m$ denotes the number of explored layers.
As $\zeta$ decreases, rEMD consistently improves for every log, reflecting better model fitting. However, this comes at the cost of longer runtimes due to the larger number of layers. For example, for the \texttt{BPIC13\_c} log, reducing $\zeta$ from $0.01$ to $0.0025$ improves rEMD from $0.082$ to $0.062$ but increases runtime from $563$ seconds to $2770$ seconds.

The results in Table~\ref{tab:exp1} were obtained with $N\!=\!100$ particles, 
using $\epsilon=99\%$ confidence level and confidence interval width $\delta\!=\! 0.1$ which, as illustrated in Section~\ref{result:hasl},  provides one with sufficiently good accuracy.

\begin{table}[t]
    \centering
    \renewcommand{\arraystretch}{1.2}
    \resizebox{\textwidth}{!}{%
    \begin{tabular}{l p{1ex} l l p{1ex} l p{2ex} l l l p{1ex} l l p{1ex} l l p{1ex} l p{1ex} l}
        \toprule 
        
        Event log & & $|Tr|$ & $|T|$ & & \multicolumn{5}{l}{ABC-SMC opt} & & \multicolumn{2}{l}{Unfolding opt \cite{DBLP:journals/corr/abs-2406-10817}} & & \multicolumn{2}{l}{Burke's WE \cite{DBLP:conf/icpm/BurkeLW20}} & & WAWE \cite{10680664} & & SLPN Miner \cite{10.1007/978-3-031-61057-8_3} \\
        \cmidrule(){6-10} \cmidrule(){12-13} \cmidrule(){15-16} \cmidrule(){18-18} \cmidrule(){20-20}
         & & & & & $\zeta$ & & $m$ & rEMD & runtime & & rEMD & runtime & & name & rEMD & & rEMD & & rEMD \\
        \hline \hline
        
        \multirow{3}{*}{BPIC13\_c} & & \multirow{3}{*}{$183$} & \multirow{3}{*}{$19$} & & $0.01$ & & $12$ & $0.082$ & $563s$ & & \multirow{3}{*}{$\mathbf{0.04}$} & \multirow{3}{*}{$603s$} & & \multirow{3}{*}{fork} & \multirow{3}{*}{$0.63$} & & \multirow{3}{*}{$0.2$} & & \multirow{3}{*}{T/O} \\
         & & & & & $0.005$ & & $17$ & $0.081$ & $1326s$ & & & & & & & \\
         & & & & & $0.0025$ & & $24$ & $0.062$ & $2770s$ & & & & & & & \\ \hline
         
        \multirow{3}{*}{BPIC13\_i} & & \multirow{3}{*}{$1511$} & \multirow{3}{*}{$19$} & & $0.01$ & & $17$ & $0.23$ & $9401s$ & & \multirow{3}{*}{$\mathbf{0.18}$} & \multirow{3}{*}{$87086s$} & & \multirow{3}{*}{lhpair} & \multirow{3}{*}{$0.7$} & & \multirow{3}{*}{$0.69$} & & \multirow{3}{*}{T/O} \\
         & & & & & $0.005$ & & $24$ & $0.19$ & $24427s$ & & & & & & \\
         & & & & & $0.0025$ & & $32$ & $\mathbf{0.18}$ & $108585s$ & & & & & & & \\ \hline
         
        \multirow{3}{*}{BPIC13\_o} & & \multirow{3}{*}{$108$} & \multirow{3}{*}{$20$} & & $0.01$ & & $6$ & $0.19$ & $61s$ & & \multirow{3}{*}{$\mathbf{0.076}$} & \multirow{3}{*}{$137s$} & & \multirow{3}{*}{pairs} & \multirow{3}{*}{$0.24$} & & \multirow{3}{*}{$0.26$} & & \multirow{3}{*}{T/O} \\
         & & & & & $0.005$ & & $15$ & $0.14$ & $340s$ & & & & & & & \\
         & & & & & $0.0025$ & & $20$ & $0.09$ & $1352s$ & & & & & & & \\ \hline
         
        \multirow{3}{*}{BPIC17\_ol} & & \multirow{3}{*}{$19$} & \multirow{3}{*}{$11$} & & $0.01$ & & $8$ & $0.09$ & $197s$ & & \multirow{3}{*}{$0.09$} & \multirow{3}{*}{$1.27s$} & & \multirow{3}{*}{freq} & \multirow{3}{*}{$0.09$} & & \multirow{3}{*}{$0.08$} & & \multirow{3}{*}{$0.25$} \\
         & & & & & $0.005$ & & $19$ & $\mathbf{0.07}$ & $434s$ & & & & & & & \\
         & & & & & $0.0025$ & & $29$ & $\mathbf{0.04}$ & $4583s$ & & & & & & & \\ \hline
        
        \multirow{3}{*}{BPIC20\_dd} & & \multirow{3}{*}{$99$} & \multirow{3}{*}{$43$} & & $0.01$ & & $23$ & $0.086$ & $2799s$ & & \multirow{3}{*}{$\mathbf{0.02}$} & \multirow{3}{*}{$3946s$} & & \multirow{3}{*}{fork} & \multirow{3}{*}{$0.93$} & & \multirow{3}{*}{$0.62$} & & \multirow{3}{*}{T/O} \\
         & & & & & $0.005$ & & $28$ & $0.082$ & $5007s$ & & & & & & & \\
         & & & & & $0.0025$ & & $29$ & $0.072$ & $4583s$ & & & & & & & \\ \hline
        
        \multirow{3}{*}{BPIC20\_rfp} & & \multirow{3}{*}{$89$} & \multirow{3}{*}{$51$} & & $0.01$ & & $14$ & $0.58$ & $2261s$ & & \multirow{3}{*}{$0.41$} & \multirow{3}{*}{$59819s$} & & \multirow{3}{*}{freq} & \multirow{3}{*}{$0.99$} & & \multirow{3}{*}{$0.98$} & & \multirow{3}{*}{T/O} \\
         & & & & & $0.005$ & & $28$ & $\mathbf{0.28}$ & $7692s$ & & & & & & & \\
         & & & & & $0.0025$ & & $76$ & $\mathbf{0.08}$ & $57634s$ & & & & & & & \\ \hline
        
        \multirow{3}{*}{Roadfines} & & \multirow{3}{*}{$231$} & \multirow{3}{*}{$34$} & & $0.01$ & & $12$ & $0.29$ & $1180s$ & & \multirow{3}{*}{$0.08$} & \multirow{3}{*}{$1276s$} & & \multirow{3}{*}{pairs} & \multirow{3}{*}{$0.27$} & & \multirow{3}{*}{$0.48$} & & \multirow{3}{*}{T/O} \\
         & & & & & $0.005$ & & $20$ & $0.11$ & $2312s$ & & & & & & & \\
         & & & & & $0.0025$ & & $24$ & $\mathbf{0.04}$ & $5201s$ & & & & & & & \\
        \bottomrule
    \end{tabular}
    }
    \caption{Comparison of rEMD distances obtained with ABC-HASL parameter inference against alternative approaches: best measured distances are in bold.}

    \label{tab:exp1}
\end{table}

ABC-SMC consistently discovers more accurate parameters (i.e., lower rEMD) than ``Weight estimation'', ``WAWE'' and ``SLPN Miner'', and delivers competitive and sometimes superior performance compared to ``Unfolding optimization''. While ``Unfolding optimization'' can occasionally yield slightly better rEMD values, ABC-SMC remains competitive, especially at finer thresholds (e.g., $\zeta=0.0025$), and is generally faster at coarser ones (e.g., $\zeta=0.01$). This trade-off between accuracy and efficiency makes ABC-SMC a practical option when both are important. It performs robustly across logs of varying complexity, from \texttt{BPIC17\_ol} to large-scale cases like \texttt{BPIC13\_i} and \texttt{BPIC20\_rfp}. For instance, on \texttt{BPIC20\_rfp} (51 transitions), ABC-SMC achieves a better rEMD (0.28) than ``Unfolding opt'' (0.41) in under one-seventh of the time. While runtime increases with log complexity, especially for logs with many unique traces like \texttt{BPIC13\_i}, ABC-SMC remains effective. This overhead stems from the extended PDF support used in \cosmos~for estimating the stochastic language, which increases computational cost.



The particle search in ABC-SMC is parallelizable, offering substantial potential to reduce runtime. All reported runtimes were obtained using 16 parallel jobs. On a supercomputer, parallelization could match the number of particles per layer, bringing ABC-SMC runtimes closer to those of other methods.

Additional experiments varying the number of particles showed that fewer particles reduced computation time without significantly affecting rEMD values. Increasing the number of particles led to longer runtimes with minimal rEMD improvement. While more particles occasionally enabled faster convergence, the accuracy gains did not offset the higher computational cost. However, using more particles yields a more precise posterior over transition weights, offering better insights into their influence on the stochastic behavior of the net.

\subsection{ABC estimates of the marginal posterior distributions}

The $n$ particles produced by the ABC procedure approximate samples from the target posterior distribution. Here we present the resulting marginal distributions over transition weights for one of the considered logs, \texttt{BPIC17\_ol}, based on the particles generated by Algorithm~\ref{alg:abc_pmc}.

\begin{figure}[t]
  \centering
  \resizebox{0.7\textwidth}{!}{%
    \begin{tikzpicture}
    \tikzstyle{place}=[circle,draw,minimum size=4mm, rounded corners=0pt]

    \node[place,tokens=1,label=left:{\footnotesize $source$}] (source) at (0,0) {};
    \node[transition,label=above:{\footnotesize $t_0$}] (t0) at (0.75,0) {$a$};
    \node[place] (p1) at (1.5,0) {};
    \node[transition,label=above:{\footnotesize $t_1$}] (t1) at (2.25,0) {$b$};
    \node[place] (p2) at (3,0) {};
    \node[transition] (t2) at (3.75,0) {$c$};
    \node[above=of t2, xshift=0.4cm, yshift=-1.2cm] {\footnotesize $t_2$};
    \node[transition] (t3) at (3.75,-0.5) {$d$};
    \node[below=of t2, xshift=0.4cm, yshift=0.8cm] {\footnotesize $t_3$};
    \node[transition, fill=black,label=above:{\footnotesize $t_4$}] (t4) at (4.5,0.65) {};
    \node[place] (p3) at (4.5,-0.25) {};
    \node[transition, fill=black] (t5) at (5.25,0) {};
    \node[above=of t5, xshift=-0.4cm, yshift=-1.2cm] {\footnotesize $t_5$};
    \node[transition] (t6) at (5.25,-0.5) {$e$};
    \node[below=of t6, xshift=-0.4cm, yshift=1.3cm] {\footnotesize $t_6$};
    \node[place] (p4) at (6,0) {};
    \node[transition] (t7) at (7,1) {$f$};
    \node[above=of t7, xshift=0.4cm, yshift=-1.2cm] {\footnotesize $t_7$};
    \node[transition] (t8) at (7,0.5) {$g$};
    \node[above=of t8, xshift=0.4cm, yshift=-1.3cm] {\footnotesize $t_8$};
    \node[transition] (t9) at (7,0) {$h$};
    \node[above=of t9, xshift=0.4cm, yshift=-1.25cm] {\footnotesize $t_9$};
    \node[transition, fill=black] (t10) at (7,-0.5) {};
    \node[above=of t10, xshift=0.45cm, yshift=-1.6cm] {\footnotesize $t_{10}$};
    \node[place, label=right:{\footnotesize $sink$}] (p5) at (8,0) {};

    \draw (source) [->] to (t0);
    \draw (t0) [->] to (p1);
    \draw (p1) [->] to (t1);
    \draw (t1) [->] to (p2);
    \draw (p2) [->, bend left=20] to (t4);
    \draw (p2) [->] to (t2);
    \draw (p2) [->, bend right=10] to (t3);
    \draw (t2) [->] to (p3);
    \draw (t3) [->] to (p3);
    \draw (p3) [->] to (t5);
    \draw (p3) [->] to (t6);
    \draw (t4) [->, bend left=15] to (p4);
    \draw (t5) [->] to (p4);
    \draw (t6) [->] to (p4);
    \draw (p4) [->, bend left=20] to (t7);
    \draw (p4) [->, bend left=10] to (t8);
    \draw (p4) [->] to (t9);
    \draw (p4) [->, bend right=20] to (t10);
    \draw (t7) [->, bend left=20] to (p5);
    \draw (t8) [->, bend left=10] to (p5);
    \draw (t9) [->] to (p5);
    \draw (t10) [->, bend right=20] to (p5);
    
\end{tikzpicture}
  }
  \caption{WN model discovered from the \texttt{BPIC17\_ol} log.}
  \label{fig:bpic17olwn}
\end{figure}
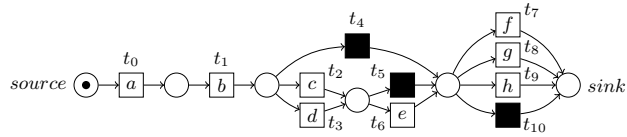

The \texttt{BPIC17\_ol} log describes a loan application process from a Dutch financial institution and involves eight distinct activities. The corresponding WN, mined using the inductive miner algorithm\footnote{Using the inductive miner algorithm.}, is shown in Figure~\ref{fig:bpic17olwn}. It consists of 11 transitions: 8 labeled with log activities and 3 silent. Since the net contains no loops, its language $L_N$ is finite.

\begin{figure}[b]
    \centering
    \begin{tabular}{cccccc}
    
        \begin{tabular}{c}
            \includegraphics[scale=0.11]{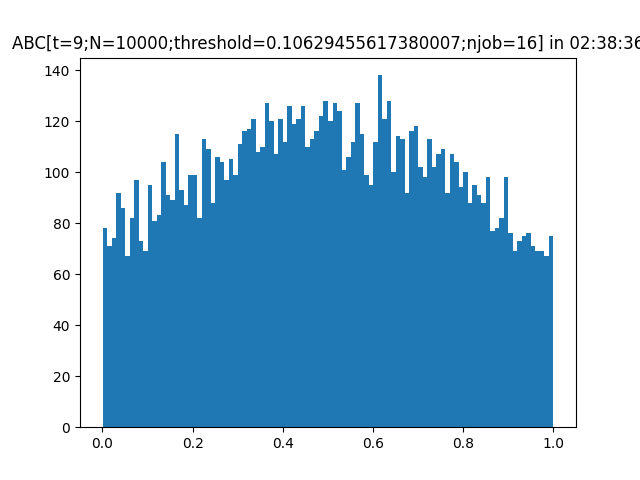} \\ 
            $\lambda(t_0) = a$ 
        \end{tabular}
        &
        \begin{tabular}{c}
            \includegraphics[scale=0.11]{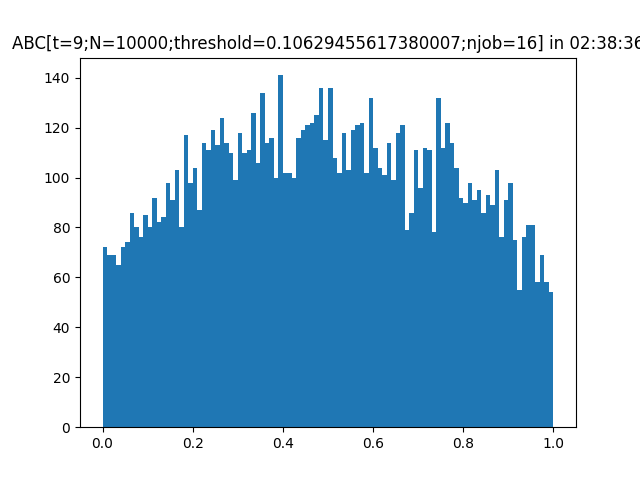} \\ 
            $\lambda(t_1) = b$ 
        \end{tabular} 
        & 
        \begin{tabular}{c}
            \includegraphics[scale=0.11]{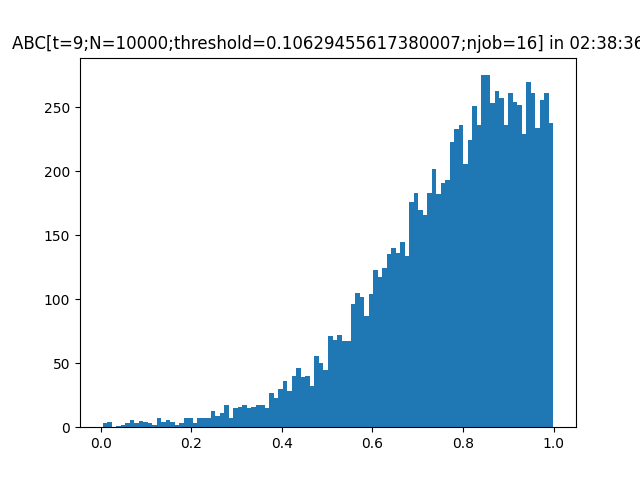} \\
            $\lambda(t_2) = c$
        \end{tabular} 
        & 
        \begin{tabular}{c}
            \includegraphics[scale=0.11]{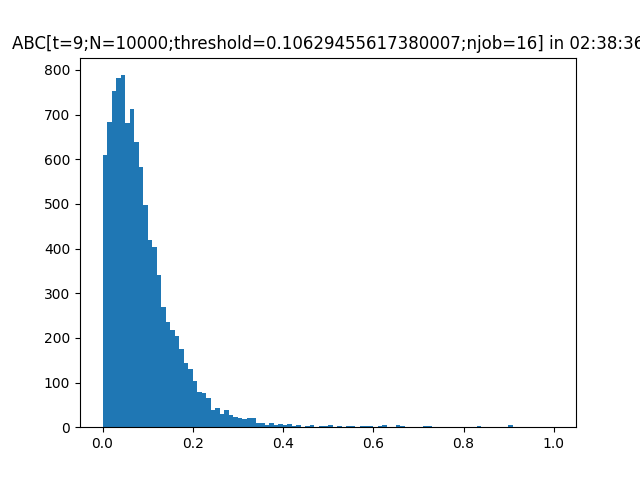}\\ 
            $\lambda(t_3) = d$
        \end{tabular}
        &
        \begin{tabular}{c}
            \includegraphics[scale=0.11]{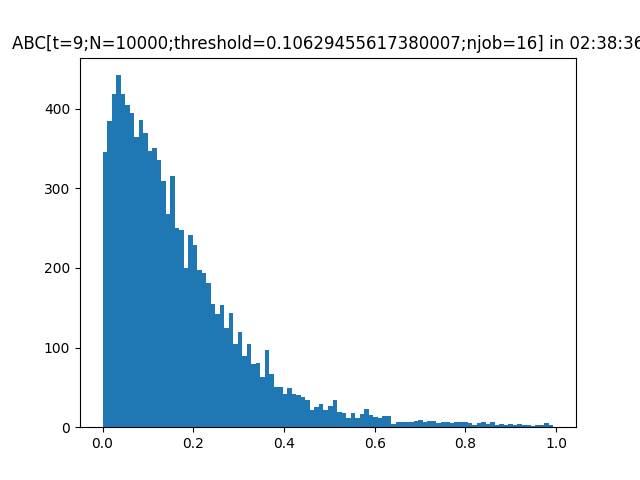} \\
            $\lambda(t_4) = \tau$
        \end{tabular} 
        &
        \begin{tabular}{c}
            \includegraphics[scale=0.11]{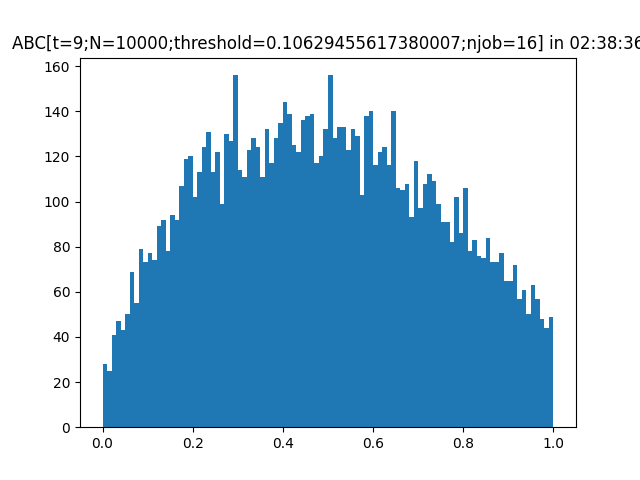} \\
            $\lambda(t_5) = \tau$
        \end{tabular} \\
        
        \begin{tabular}{c}
            \includegraphics[scale=0.11]{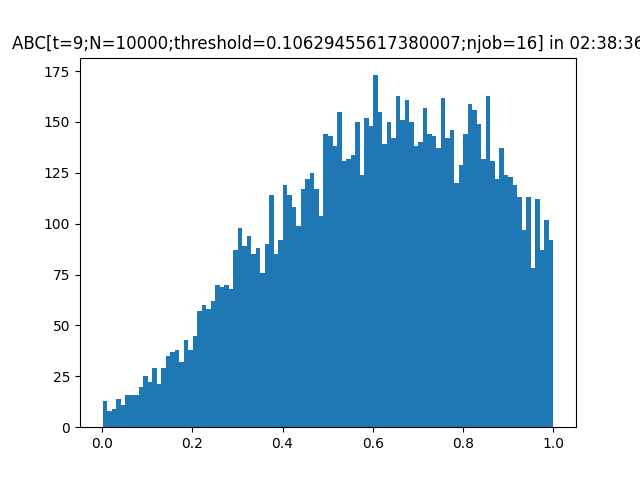} \\
            $\lambda(t_6) = e$ 
        \end{tabular}
        &
        \begin{tabular}{c}
            \includegraphics[scale=0.11]{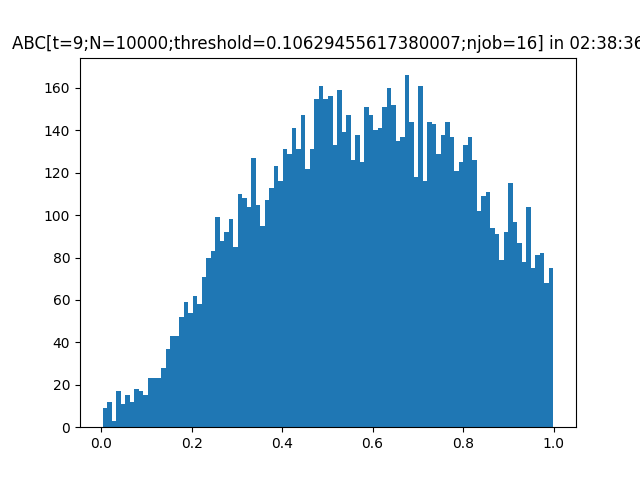} \\
            $\lambda(t_7) = f$
        \end{tabular}
        &
        \begin{tabular}{c}
            \includegraphics[scale=0.11]{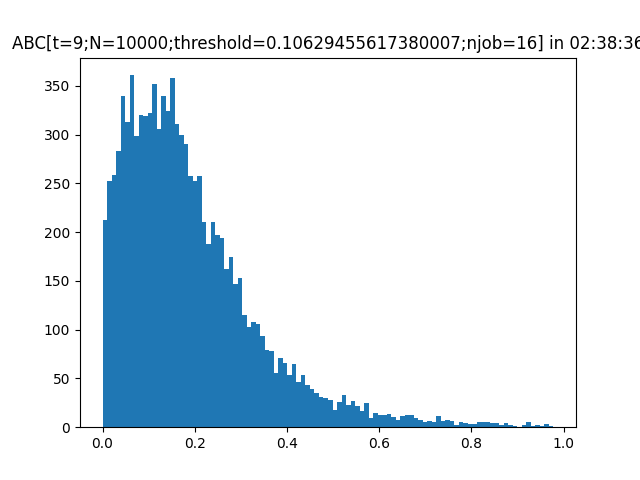}\\
            $\lambda(t_8) = g$ 
        \end{tabular}
        &
        \begin{tabular}{c}
            \includegraphics[scale=0.11]{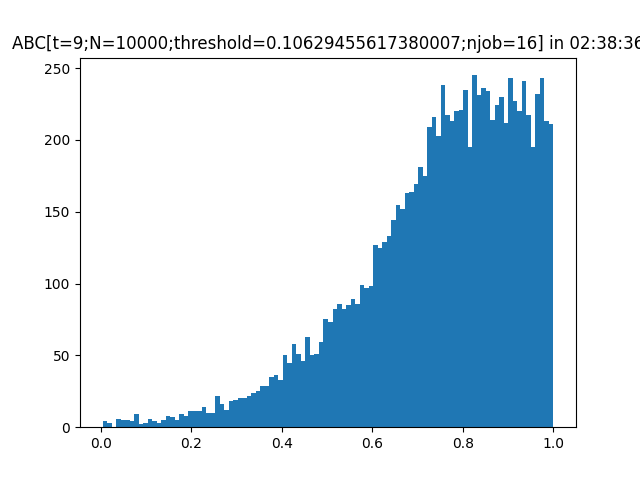} \\
            $\lambda(t_9) = h$
        \end{tabular}
        &
        \begin{tabular}{c}
            \includegraphics[scale=0.11]{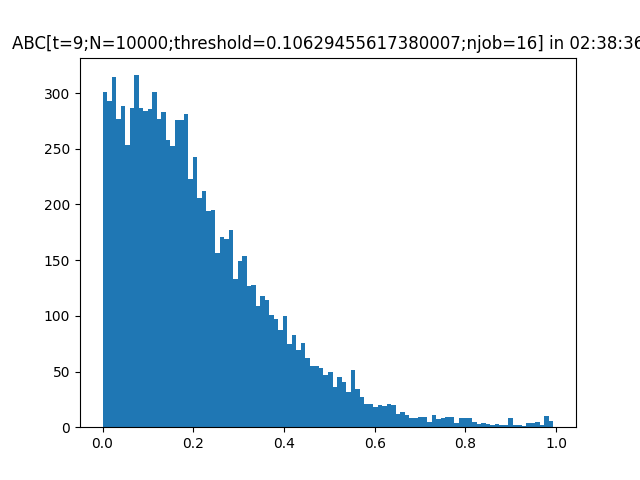} \\
            $\lambda(t_{10}) = \tau$
        \end{tabular}
        
    \end{tabular}
    \caption{Marginal posterior distributions of the weights for the 11 transitions of the BPIC17\_ol model: obtained with Algorithm~\ref{alg:abc_pmc} using $n=10000$ particles.}
    \label{fig:bpic17posteriors}
\end{figure}


Figure~\ref{fig:bpic17posteriors} illustrates the approximation of the marginal posterior distributions for the weights of the 11 transitions $t_i$ ($0\leq i\leq 10$) of the sWN (Figure~\ref{fig:bpic17olwn}) corresponding to log \texttt{BPIC17\_ol} resulting from $n=10000$ particles sampled through Algorithm~\ref{alg:abc_pmc} and using the interval $[0,1]$ as support for each weight.

The non-informative (uniform-like) distribution for certain transitions weights (i.e., $t_0$, $t_1$) indicates their little relevance w.r.t. the conformance between the net and the log stochastic languages. Conversely, transitions whose marginal concentrates the probability mass close to the supremum of the support ``dominate'' those whose probability mass is close to the infimum. For example, $t_2$ dominates both $t_3$ and $t_4$ with which it is in conflict and, similarly, $t_9$ and, to a less extent, $t_7$ dominate over $t_8$ and $t_{10}$ which they are in conflict with.

These dominance relationships reflect statistical characteristics of the 16 traces in the \texttt{BPIC17\_ol} log. For instance, the conflict among transitions $t_2$ (labeled $c$), $t_3$ (labeled $d$), and $t_4$ (silent) arises because some traces contain only $c$, others only $d$, some neither, and none both. In the log traces with $c$ occur with probability $0.924$, those with $d$ with $0.047$, and those with neither $c$ nor $d$ with $0.029$, summing to 1. This indicates that activity $c$ is far more likely than $d$ or neither. In the sWN, this distribution must be reflected in the transition weights: $t_2$ should have a much higher probability than $t_3$ and $t_4$. The marginal posterior distributions confirm this: all three are approximately a truncated Gaussian, with $t_2$ centered near 1 and $t_3$, $t_4$ near 0. This shows that $t_2$ must carry significantly more weight to accurately capture the observed trace probabilities. A similar pattern emerges in the conflict between $t_7$, $t_8$, $t_9$, and $t_{10}$. Here, $t_9$ labeled $h$, which is the most probable event in the log, has a weight close to $1$ to reflect this feature of the log.


\section{Conclusion}
\label{sec:conclusion}

We introduced a novel,  simulation-based framework for stochastic process discovery. The framework replaces the computationally heavy, exact computation of the discovered model's stochastic language  (of~\cite{cry2024framework,10.1007/978-3-031-61057-8_3}) by an arbitrarily precise approximation obtained via synchronization with a stochastic language detector automaton. Optimal model parameters are obtained via an adaptation of  \emph{likelihood free} Bayesian inference scheme where the stochastic language approximation engine is plugged in. 
Through experiments on real-life event logs, we have demonstrated that, in terms of conformance, our approach 1) outperforms the weight estimators \cite{DBLP:conf/icpm/BurkeLW20} 
as well as the WAWE framework,  2) it manages to optimize models that cannot be treated with the SLPN tool~\cite{10.1007/978-3-031-61057-8_3} and 3) it may result in a better conformance w.r.t. unfolding based optimisation~\cite{DBLP:journals/corr/abs-2406-10817}. Therefore we argue  that simulation-based discovery of optimal models is a valuable alternative for scaling to real-world process mining applications. Finally it is worth  pointing out that   posterior distribution analysis (e.g., Fig.~\ref{fig:bpic17posteriors}), supported by our framework (but not by alternative ones), provides the user with useful added insights. 
We envisage two directions as future developments: first the integration, within the framework, of entropy relevance~\cite{ALKHAMMASH2022101922} as an alternative to rEMD based objective function as this could reduce the runtime of optimization and secondly  to consider the extension to timed stochastic models.

\bibliographystyle{plain} 
\bibliography{biblio}

\appendix
\section{Appendix}
\label{appendix}

\subsection{HASL model checking}
\label{sec:hasl}

The HASL statistical model checking  framework (Figure~\ref{fig:hasl}) allows for assessing a GSPN model $N_s$, via  a property $\varphi\equiv({\cal A},Z)$  formally encoded by a combination of  linear hybrid automaton ${\cal A}$ and a target expression $Z$. The functioning of the framework can be summarised as follows:  a sufficiently large number of (finite) traces of model  $N_s$ are sampled (via simulation) and synchronised  (\emph{on-the-fly}) with the LHA ${\cal A}$ and those that meet the acceptance condition(s) of ${\cal A}$ are used (together with the statistics collected in the variables of ${\cal A}$)  to build  a $\epsilon$\%-confidence interval estimate (with width $\delta$)  of the  measure of interest $Z$, which (in the reminder we use $\mathit{CI}(Z, \mathcal{A}\times N_s, \epsilon, \delta)$ to denote such confidence interval). 

\begin{figure}[h]
    \centering
    \includegraphics[scale=0.3]{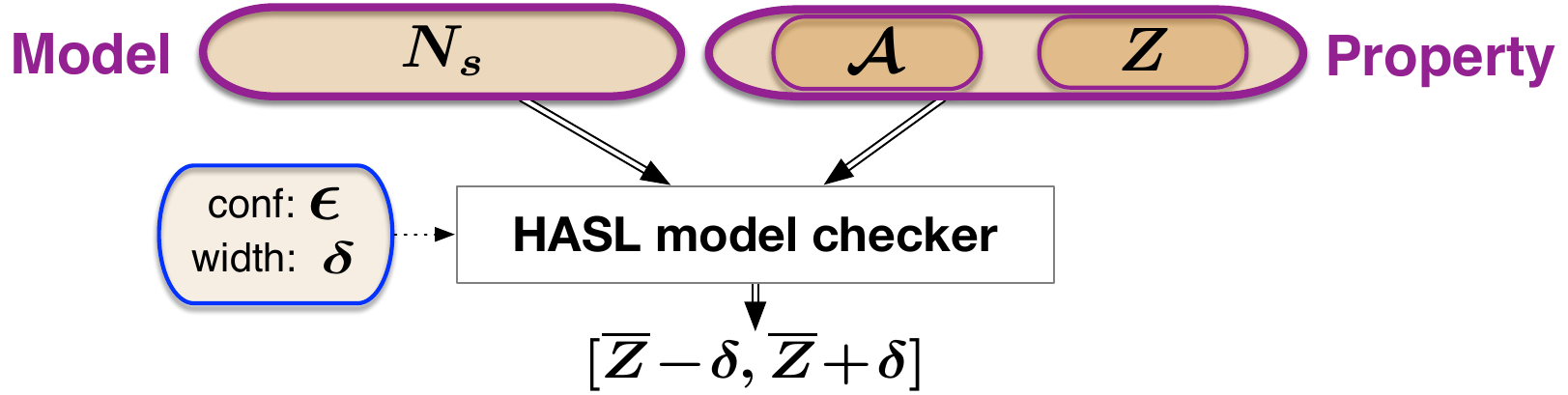}
    \caption{The HASL statistical model checking scheme}
    \label{fig:hasl}
\end{figure}

The synchronization of a GSPN and corresponding LHA is formalised by the definition of the product process $N_s\!\times\!\aut$ whose formal semantics (see~\cite{BALLARINI201553})) leads to the stochastic simulation procedure  implemented by the HASL model checker. {\bf Synchronization of a GSPN model with a LHA}. 
For the sake of space, here, we only provide an intuitive (informal) description of such synchronization based on the Example~\ref{ex:lha}. The LHA in    Figure~\ref{fig:lhaexample} (right) is designed for synchronisation with the GSPN model $N_s$ Figure~\ref{fig:lhaexample} (left). Synchronisation starts in the initial state $(source,l_{start},[0,0,0])$ of product process $N_s\times {\cal A}_{N_s}$, i.e. with $source$ being the initial marking of $N_s$, $l_{start}$ the initial location of ${\cal A}_{N_s}$ and $[0,0,0]$ the initial values for the 3 variables $(x_1,x_2,x_3)$ of  ${\cal A}_{N_s}$ and terminates as soon as no transitions of the product process $N_s\times {\cal A}_{N_s}$ is enabled (trace rejected), or as soon as the final location  $l_{end}$ of ${\cal A}_{N_s}$ is reached (trace accepted).    
As the acceptance condition of  ${\mathcal A_{N_s}}$ is given by the guard $x_1\!\geq \!T$ of the \emph{autonomous} (i.e. $\sharp$) edge $l_{start}\to l_{end}$, it follows that  ${\mathcal A_{N_s}}$ accepts all (timed) traces of $N_s$ as soon as they reach time at least $T$. Before reaching time $T$ a simulated trace will trigger traversal of  the upper (resp. lower) synchronised $l_{start}\to l_{start}$ self-loop edge  on every occurrence of the $N_s$ transition labeled \texttt{b} (resp. any other transition but that labeled \texttt{b}). Notice that variable $x_3$ is used as a counter of the occurrences of the $N_s$ transition \texttt{b}, as it is incremented on traversal of the corresponding  synchronised $l_{start}\to l_{start}$ edge. Finally, on reaching the acceptance location $l_{end}$ (as soon as the simulation time is $x_1\!\geq \! T$, with $T$ being a constant of the LHA, and on condition $N_s$ has a token in place $sink$), variable $x_2$ is divided by $T$ which corresponds to the average number of combined tokens   in places $p_2$ and $p_5$ observed along a simulation that lasted at least $T$ time units and ended with  place $sink$ containing a token.

\medskip
\noindent
{\bf HASL target expression.} The second component of an HASL formula $\varphi\equiv({\cal A},Z)$ is an expression $Z$ resulting from  a grammar\footnote{see (\ref{haslexp}) in Section~\ref{appendix:haslexp}}  over the  variables of the considered LHA~\cite{BALLARINI201553}. In practice  $Z$ is obtained by application of some  \emph{statistical} operator (i.e. $PROB()$, $AVG(\cdot)$,  $PDF(\cdot,\cdot,\cdot,\cdot)$, $CDF(\cdot,\cdot,\cdot,\cdot)$) on (algebraic combinations) of LHA variables.  For the sake of  simplicity here we present a few instances of Z expressions   referred to the LHA 
and corresponding GSPN model 
of Example~\ref{ex:lha}. 

\begin{itemize}
    \item $Z_1=\mathit{AVG}(last(x_2))$ represents the \emph{average number of tokens contained in places $p_2$ and $p_5$ within time at least $T$} 
    \item $Z_2=\mathit{PDF}(last(x_2),0.01,0,2)$ represents the PDF of the \emph{average  number of tokens contained in places $p_2$ and $p_5$ within time at least $T$} approximated using $[0,2]$ as support set and discretizing $[0,2]$ with buckets of width $0.01$
    \item $Z_3=\mathit{AVG}(min(x_3))$ represents the \emph{minimum number of occurrences of transition \texttt{b}  within time at least  $T$} 
    
\end{itemize}

In the context of  the HASL-based framework for optimised process discovery we present in Section~\ref{sec:method} we  only refer to one kind of HASL expression, i.e. that based on the $PDF(\cdot,\cdot,\cdot,\cdot)$ operator. 

\noindent
\textit{Remark.} HASL has been proved a very rich property language which allows to combine time-constraints and arbitrarily complex (\emph{signal-analysis like}) statistics  to characterise the  behavior of interest. We briefly illustrated its full semantics referring to an actual (timed) GSPN model, although in the context of this paper we will apply HASL to analyse sWNs, i.e. an untimed subclass of GSPNs.

\subsection{Grammar of HASL target expressions}
\label{appendix:haslexp}
A HASL target expression $Z$ built on top of  a  LHA  $\mathcal A_{N_s}\!=\!\langle E, L,  I, F, X, \flow, \Lambda, \rightarrow \rangle$   is  defined by the following grammar: 

\begin{footnotesize}
  \begin{equation}
  \label{haslexp}
  \begin{split}
 Z  & ::=  \ AVG(Y)\ |\ Z+Z\ |\ Z \times Z \ |\ Pdist\\
 Pdist  & ::=  \ PDF(Y,step,start,stop))\ |\  {CDF(Y,step,start,stop)}\ |\  {PROB()}\\
 Y  & ::= \ c\ |\ Y+Y\ |\ Y \times Y\ |\ Y/Y\ |\ {last(y)}\ |\ min(y)\  \  |\ {max(y)}\\
 y  & ::= \ c\ |\ x\ |\ y+y\ |\ y \times y\ |\ y/y    
  \end{split}
  \end{equation}
\end{footnotesize}  
where: 
\begin{itemize}
    \item $y\in X$ is a variable of $\mathcal A_{N_s}$. 
    \item $c\in \mathbb{R}$ is a constant. 
    \item $last(y)$ is the value that $y$ has on the accepted trace  at the end of synchronisation.
    \item $avg(y)$ is the mean value measured for $y$ along the accepted trace.
    \item $min(y)$ is the minimum value that  $y$ has taken along  the accepted trace.
    \item $max(y)$ is the maximum value that  $y$ has taken along  the accepted trace.
    \item $PROB()$ is the ratio of accepted traces over the total number of  traces simulated
    \item $AVG(Y)$ is the mean value (i.e. the center of the confidence interval) of $Y$. 
    \item $PDF(Y,step,start,stop))$ is the approximation of the probability distribution function for $Y$, resulting by using $[start,stop]$ as discretised support for the distribution (i.e. $[start,stop]$ is split in $(stop-start)/step$ equally sized buckets of length $step$.
    \item $CDF(Y,step,start,stop))$ is the approximation of the cumulative distribution function for $Y$, resulting by using $[start,stop]$ as discretised support for the distribution (i.e. $[start,stop]$ is split in $(stop-start)/step$ equally sized buckets of length $step$.
\end{itemize}

\subsection{Approximate Bayesian Computation\label{app:ABC}} 

Approximate Bayesian Computation (ABC)~\cite{Marin2011,Sisson2018}  methods are concerned with estimating the posterior distribution of a model's parameters $\theta$ based on some observed (experimental) data $y_{exp}$. Considering a prior distribution on the parameters $\pi(.)$, observations $y_{exp} \in \mathcal{Y}$ and likelihood function $p(.|\theta)$ of a model, the objective of Bayesian estimation is to determine the posterior distribution:

\begin{footnotesize}
\begin{equation}
    \pi(\theta|y_{exp}) = \frac{p(y_{exp}|\theta) \pi(\theta)}{\int_{\theta'} p(y_{exp}|\theta') \pi(\theta')\, d\theta'}
\end{equation}
\end{footnotesize}
\noindent
In most models, the likelihood functions $p(y_{exp}|\theta)$ is too expensive to compute or even intractable, hindering the determination of the posterior distribution. Therefore ABC algorithms allow one to obtain an arbitrarily precise approximation, denoted $\pi_{\mathit{ABC},\epsilon}$ ($\epsilon\in\mathbb{R}^*_+$ being a tolerance value), of the posterior $\pi(\theta|y_{\mathit{exp}})$. In its simplest, \emph{rejection sampling}, form (Algorithm~\ref{alg:abc}).  ABC consists of a simple iterative procedure where $n$ parameters value $\theta'$ are selected from the parameter space through sampling from  the prior distribution $\pi()$, up until the distance between the traces issued by the corresponding model instance ($y'\sim p(.|\theta'))$)  and the observations $y_{exp}$  is below tolerance $\epsilon$ (i.e. $\rho(\eta(y'),\eta(y_{exp}))\leq \epsilon$~\footnote{where $\eta: \mathcal{Y} \rightarrow \mathcal{S}\subset\mathbb{R}^{k_1}$ is a function that computes summary statistics on the observations and  $\rho : \mathcal{S} \times \mathcal{S} \rightarrow \mathbb{R}^+$ is a distance in the space of summary statistics.} ). 
Though effective, ABC rejection sampling suffers of slow convergence, particularly for small tolerance values $\epsilon$. Therefore, the sequential Monte Carlo extension of ABC, named ABC-SMC~\cite{Beaumont2008}, has been introduced to speed up the search of acceptable parameters, essentially by means of a multi-level, parameter search procedure through which parameters are progressively accepted using decreasing tolerance levels. 
\vskip -.5cm
\begin{algorithm}[H]
\scriptsize
\caption{ABC rejection sampling}\label{alg:abc}
\begin{algorithmic}
    \Require $y_{exp}$ (observations), $\epsilon$ (tolerance), $\rho$ (distance metric), $\eta$ (summary statistics)
    \Ensure $(\theta_i)_{0\leq i \leq n}$ drawn from $\pi_{ABC,\epsilon}$
    \For {$i = 1:n$}
    \Repeat
    \State $\theta ' \sim \pi(.)$
    \State $y' \sim p(.|\theta ')$
    \Until {$\rho(\eta(y'), \eta(y_{exp})) \leq \epsilon$}
    \State $\theta_i \gets \theta'$
    \EndFor
\end{algorithmic}
\end{algorithm}
\vskip -.5cm
\noindent 

Since ABC algorithms rely on a distance measure ($\rho$) it has been shown they can be adapted to calibrating models w.r.t. specific behavioral characteristics  (rather than w.r.t. observations $y_{exp}$) as long as an adequate distance can be defined to measure how far a model instance (issued by a parameter vector $\theta_i$) is from exhibiting the desired characteristic (e.g. see~\cite{bentriou2019,DBLP:journals/tcs/BentriouBC21,DBLP:conf/cmsb/BallariniBC23}). 

In Section~\ref{sec:method}, we introduce a novel version of the ABC-SMC algorithm which, based on the restricted Earth Movers Distance (rEMD), allows for inferring the weight's parameters of a sWN model $N_s$ (mined from a given event log $E$) so to minimize the distance between the stochastic language $L_{N_s}$, issued by the sWN, and that of the event log $L_E$.

\subsection{Functioning of the ABC Sequential Monte Carlo}
\label{appendix:abcsbc}

\begin{figure}
    \centering
    \includegraphics[scale=.5]{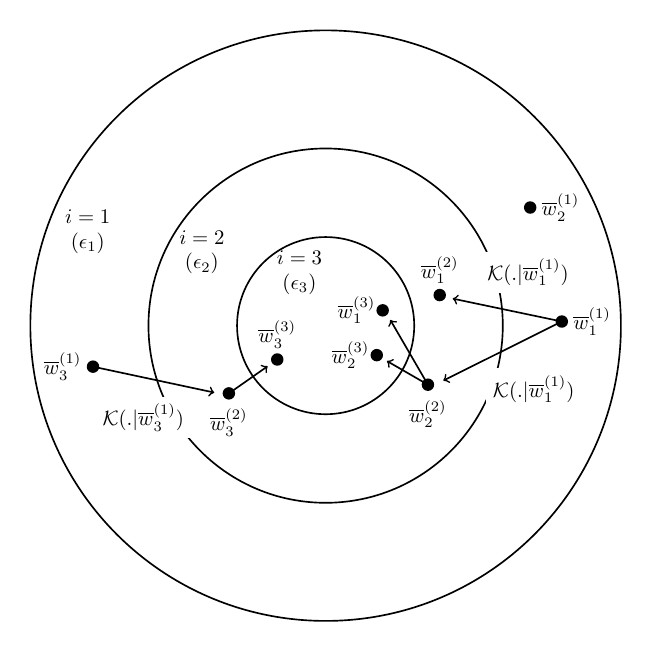}
    \caption{Functioning of the ABC Sequential Monte Carlo  scheme with $m=3$ layers and $n=3$ particles.}
    \label{fig:smcabc}
\end{figure}

Figure~\ref{fig:smcabc} illustrates the functioning of Algorithm~\ref{alg:abc_pmc} in a toy configuration consisting of $m=3$ layers and $n=3$ particles per layer. Each distinct layer (circle) represents an iteration where particles are propagated, weighted, resampled, and accepted according to a corresponding tolerance $\epsilon_i$ ($1\leq i\leq 3$). Accepted particles are shown as distinct points within each layer, while arrows depict particle propagation from one layer to the next. These transitions represent the particle propagation through three steps: (1) selection of the next particle to propagate, based on its weight $\delta_j$; (2) movement of the selected particle $\overline{w}_j^{(i)}$ via sampling from a kernel $K(.|\overline{w}_j^{(i)})$ centered around it; and (3) re-weighting of the particles based on their likelihood. At the first layer ($i=1$), the initial particles $\{\overline{w}^{(1)}_1, \overline{w}^{(1)}_2, \overline{w}^{(1)}_3\}$ are sampled from the prior distribution $\mathcal{U}(0,1)^3$, and their initial weights are set to $\delta_1\!=\!\delta_2\!=\!\delta_3\!=\!1/3$. The particles are propagated to the next layer ($i=2$). For example, in Figure~\ref{fig:smcabc}, particle $\overline{w}_1^{(1)}$ is selected once, particle $\overline{w}_3^{(1)}$ is selected twice, and particle $\overline{w}_2^{(1)}$ is never selected. The selected particles are then moved by sampling from a kernel centered on them.

\end{document}